\documentclass[aps,prd,preprintnumbers,nofootinbib,floatfix,epsfig]{revtex4-1}
\usepackage{dcolumn}
\usepackage{graphicx}
\usepackage{tikz}
\usepackage{color}
\usepackage{amsmath}
\usepackage{amssymb}
\usepackage{amsthm}
\usepackage{latexsym}
\usepackage[hmargin=1.8cm,vmargin=1.8cm]{geometry}
\usepackage{bm}
\usepackage{multirow}
\usepackage{cancel}
\usepackage{hyperref}
\usepackage{tabularx}
\usepackage{float}
\usepackage{subcaption}
\restylefloat{table}

\def \tautau {\tau^{+}\tau^{-}}

\def\pb1{\,\text{pb}^{-1}}
\def\fb1{\,\text{fb}^{-1}}
\def\ab1{\,\text{ab}^{-1}}

\begin{document}
\vspace*{2cm}
\title{Implications of  light charged Higgs boson at the LHC Run III in the 2HDM}

\author{Abdesslam Arhrib}
\email[]{aarhrib@gmail.com}
\affiliation{\small Facult\'e des Sciences et Techniques, Abdelmalek Essaadi University, B.P. 416, Tangier, Morocco}

\author{Rachid Benbrik}
\email[]{r.benbrik@uca.ac.ma}
\affiliation{\small Laboratoire de Physique Fondamentale et Appliquée Safi, Faculté Polydisciplinaire de Safi, Safi, Morocco.}
\author{Hicham Harouiz}
\email[]{r.benbrik@uca.ac.ma}
\affiliation{\small Laboratoire de Physique Fondamentale et Appliquée Safi, Faculté Polydisciplinaire de Safi, Safi, Morocco. }

\author{Stefano Moretti}
\email[]{s.moretti@soton.ac.uk}
\affiliation{\small School of Physics and Astronomy, University of Southampton,\\
    Southampton, SO17 1BJ, UK}

\author{ Yan Wang}
\email[]{wangyan@imnu.edu.cn}
\affiliation{\small 
College of Physics and Electronic Information, Inner Mongolia Normal University, Hohhot 010022, PR China}

\author{Qi-Shu Yan}
\email[]{yanqishu@ucas.ac.cn}
\affiliation{\small 
School of Physics Sciences, University of Chinese 
Academy of Sciences, Beijing 100039, PR China}
\affiliation{\small 
Center for future high energy physics,  Chinese  
Academy of Sciences, Beijing 100039, PR China}

\date{\today}

\vspace*{-3cm}

\begin{abstract}
In this study, we focus on the bosonic decays of light charged Higgs boson (i.e., with $M_{H^\pm}<m_t$)
 in the 2-Higgs Doublet Model (2HDM) Type-I. To study the signal of such a charged Higgs state at the Large Hadron Collider (LHC), in a scenario where the $H^0$ boson is the Standard Model (SM)-like one already discovered, 
we assume that it decays mainly via $h^0W^{\pm *}$  and/or  $H^\pm\to A^0W^{\pm *}$ (i.e., via an off-shell $W^{\pm}$ boson), which can reach a sizable Branching Ratio (BR) for $\tan\beta\geq4$, when the exclusion bounds from $H^\pm\to\tau\nu$ and $c{s}$  searches get weaker. By using six Benchmark Points (BPs), which are consistent with current LHC constraints, we perform a Monte Carlo (MC) study and examine the sensitivity of the LHC to light charged Higgs boson decaying via the above bosonic modes and produced in top decay following both single top and top pair production processes. Our findings demonstrate that, when the integrated luminosity can reach 100 fb$^{-1}$, the LHC has the potential to either discover or rule out most of these BPs via either of these two production and decay channels or both.
\end{abstract}

\maketitle

\section{Introduction} 

Following the discovery of a 125 GeV Higgs boson in the first run of the LHC \cite{bib1,bib2},  several studies of its properties have now been undertaken. These cover the measurement of its mass, width, spin, CP properties as well as couplings. The current situation is that the measured Higgs signal rates in all available channels agree with the SM predictions at the $\sim 2\sigma$ level \cite{bib3}. However, the possibility that the observed Higgs boson state (hereafter, referred to as $H_{\rm obs}$)
could belong to a model with an extended Higgs sector, such as the SM with an extra singlet, doublet and/or triplet, has not been ruled out.  Amongst such higher Higgs representations,  those with an extra doublet or triplet
also contain one or more charged Higgs bosons in their (pseudo)scalar spectrum.  The discovery of such charged Higgs bosons would then be an eminent signal of a
multiplet Higgs sector and thus clear evidence of physics Beyond the SM (BSM). However, Nature has so far indicated that the Higgs mechanism relies upon a doublet structure, so we focus here on a BSM scenario that only employs this particular multiplet. 

One of the simplest extensions of the SM within this kind is the 2HDM, which contains two Higgs doublets, $\Phi_1$ and $\Phi_2$, used to give mass to all fermions. The particle spectrum of the 2HDM is as follows: two CP-even ($h^0$ and $H^0$, with $M_{h^0}<M_{H^0}$), one CP-odd ($A^0$) and a pair of charged ($H^\pm$) Higgs bosons. At hadron colliders, a charged Higgs boson can be produced through several channels.  Light charged Higgs states, i.e, with $M_{H^0}^\pm\leq m_t-m_b$, are copiously induced by $t\bar{t}$ production followed by the top decay $t\to bH^+$ (or the equivalent antitop mode). When kinematically allowed, $pp\to t\bar{t}\to b bH^-W^++ c.c$. provides the most important source of light charged Higgs bosons, above and beyond the yield of various direct production modes: $gb\to tH^+$ (or $gg\to t\bar{b}H^+$)  \cite{bib4}, $gg\to W^\pm H^\mp$ and $b\bar{b}\to W^\pm H^\mp$  \cite{bib5}, $q\bar{q}^\prime\to \phi H^\pm$ where $\phi$ denotes one of the three neutral Higgs bosons  \cite{bib6}, $gg\to H^+H^-$ and $q\bar{q}\to H^+H^-$  \cite{bib7}, $qb\to q^\prime H^+b$  \cite{bib8} and $c\bar{s}$, $c\bar{b}\to H^+$  \cite{bib9}. (See also Refs.  \cite{bib10,bib11} for a review of all available $H^\pm$ hadro-production modes in a 2HDM.)

Assuming a light charged Higgs boson, i.e., such that $M_{H^\pm}<m_t-m_b$, 
the ATLAS and CMS experiments have already drawn an exclusion on ${\rm BR}(t\to H^+b)\times {\rm BR}(H^\pm\to \tau\nu)$ based on the search for the corresponding decay chain  \cite{cms,atlas}. Other channels, such as $H^+\to c\bar{s}$, have also been searched for by ATLAS and CMS  \cite{bib12,bib13}. Assuming that ${\rm BR}(H^+\to c\bar{s})=100\%$,  one can set a limit on ${\rm BR}(t\to H^+b)$ to be in the range $5\%$ to $1\%$ for a charged Higgs boson mass between 90 and 150 GeV. We recall here that charged Higgs bosons have been also searched for at LEP-II using charged Higgs boson pair production followed by either $H^\pm\to \tau\nu$, $H^\pm\to cs$ or $H^\pm\to W^\pm A$  \cite{bib13}.  If the charged Higgs boson decays dominantly to $\tau\nu$
or $cs$, the LEP-II lower bound on the mass is of the order of 80 GeV while in the case where charged Higgs decay is dominated by $W^{\pm}A$, via a light CP-odd Higgs state ($M_{A^0}\sim$ 12 GeV), the lower bound on the charged Higgs mass is about 72 GeV  \cite{bib13}. 

The aim of this paper is  to show that the bosonic decays of a {\sl light} (i.e., with 72 GeV $< M_{H^\pm}< m_t$) 
charged Higgs boson, specifically, $H^\pm \to W^{\pm *}h$ and/or  
$H^\pm \to W^{\pm *}A$, in a scenario where $H^0$ is the discovered SM-like Higgs state (i.e., $H=H_{\rm obs}$), 
could be substantial and may compete with the aforementioned fermionic modes over specific regions of the 2HDM Type-I parameter space.
Our study builds upon the results of Ref. 
\cite{Arhrib:2016wpw}, yet we surpass this paper in several directions.
Firstly, we allow for $t\to b H^+$ decays in the presence of both single and double top-quark production (whereas Ref.  \cite{Arhrib:2016wpw}  only considered the latter).
Secondly, unlike that reference, which only performed an inclusive analysis, we proceed here to a full MC simulation in presence of Parton shower, Hadronisation, heavy flavor decays, and detector effects.
Thirdly, we allow here for $A^0$ decays into $\tau^+\tau^-$ pairs wherein the latter can in turn decay fully hadronically, fully leptonically and semi-leptonically (or semi-hadronically), whereas Ref. 
  \cite{Arhrib:2016wpw} made no assumption on the $A^0$ decay patterns. 

The layout of the paper is as follows. In the next section, we describe the 2HDM realization we are interested in (i.e., a Type-I)  whereas in the following one we outline the theoretical and experimental constraints acting upon it. Then we define the BPs to be used for the MC analysis, which is then described in detail. Following the discussion of the results, we will finally conclude.

\section{The 2HDM Type-I}
The 2HDM consists of two complex ${\rm SU}(2)_L$ (where $L$ indicates the isospin)  
(pseudo)scalar Higgs doublets $\Phi_{1,2}$ with the same hypercharge $Y_{1,2}=+1/2$ that give masses to SM gauge bosons  as well as fermions.  Explicitly, 
$\Phi_{1}$ and $\Phi_{2}$ are defined as
\begin{equation}
\Phi_{1}=\begin{pmatrix}
  \phi_1^{+}    \\
  (v_1+\phi^{0}_1+i\chi_1^{0})/\sqrt{2}
\end{pmatrix}\,, \quad\text{and}\quad
\Phi_{2}=\begin{pmatrix}
  \phi_2^{+}    \\
  (v_2+\phi^{0}_2+i\chi_2^{0})/\sqrt{2}
\end{pmatrix}\,.
\end{equation}
The 2HDM Lagrangian  involving the two Higgs doublets 
$\Phi_{1}$ and $\Phi_{2}$ can be written as:
\begin{equation}\label{equ:2HDM_Lagrangian}
\mathcal{L}=\sum_i |D_{\mu} \Phi_i|^2 - V(\Phi_1, \Phi_2) + \mathcal{L}_{\rm Yukawa}.
\end{equation}
The first term  is the kinetic one for the (pseudo)scalar fields that generate the gauge boson masses as well as the Higgs boson interactions with the 
gauge bosons themselves. The second term is the scalar potential and the third one is the Yukawa Lagrangian.
The covariant derivative is given by $D_\mu =\partial_\mu + i g \vec{T}_a \vec{W}_\mu^a + i g' \frac{Y_i}{2} B_\mu$,
where $W_\mu^a$ and $B_\mu$ are respectively the $SU(2)_L$ and $U(1)_Y$ gauge fields, with $g$ and  $g'$ the associated gauge coupling constants.

The most general scalar potential, of dimension-4, that is $SU(2)_L\times U(1)_Y$ invariant, CP-conserving and which possess a $Z_2$  symmetry  ($\Phi_1\to -\Phi_1$ or $\Phi_2\to -\Phi_2$) which is introduced to avoid Flavour Changing Neutral Currents (FCNCs) yet it is  softly broken (by a dimension-2 term, the one $\propto m_{12}^2$ below) to enable a non-trivial dynamics,  
is given by:
\begin{eqnarray}
\label{eq:2HDM_Higgs_potential}
 V(\Phi_1, \Phi_2) &=& m_{11}^2\Phi_1^\dag \Phi_1 + m_{22}^2\Phi_2^\dag \Phi_2 -m_{12}^2(\Phi_1^\dag \Phi_2+ H.c.) + \frac{\lambda_1}{2}(\Phi_1^\dag \Phi_1)^2 + \frac{\lambda_2}{2}(\Phi_2^\dag \Phi_2)^2  \notag \\
 & &+ \lambda_3(\Phi_1^\dag \Phi_1)(\Phi_2^\dag \Phi_2)+\lambda_4(\Phi_1^\dag \Phi_2)(\Phi_2^\dag \Phi_1)+\frac{\lambda_5}{2}   \Big[ (\Phi_1^\dag \Phi_2)^2 + H.c.\Big]\,,
\end{eqnarray}
where all parameters are real valued. After spontaneous Electro-Weak Symmetry Breaking (EWSB) of $SU(2)_L\times U(1)_Y$ down to $U(1)_{EM}$ (where $L$ indicates the Electro-Magnetic (EM) group), each doublet obtains a Vacuum  Expectation Value (VEV)  $v_i\ (i=1,2)$, such that  
they can be fixed from the EW scale through  the relation $v=\sqrt{v_1^2+v_2^2} = 246\ {\rm GeV}$. Furthermore, upon EWSB, three Goldstone bosons ($G^{\pm}$ and $G$) are absorbed as the longitudinal component of the gauge bosons ($W^\pm$ and $Z$, respectively), so that the latter gain their mass.  The remaining five degrees of freedom (out of the initian eight pertaining to the two complex doublets) become the physical Higgs bosons, namely, the aforementioned $h,H,A$ and $H^\pm$. 

From the two minimisation conditions of the 2HDM, one can 
eliminate $m_{11}^2$ and $m_{22}^2$ in favour of other (pseudo)scalar inputs, so that we are then left with  the following seven free independent real  parameters
\begin{equation}
\label{eq:Non_physical_basis}
\tan\beta(\equiv\frac{v_2}{v_1}) \,,\ m^2_{12}\,,\ \lambda_1\,,\ \lambda_2\,,\ \lambda_3\,,\ 
\lambda_4\,,\ \lambda_5.
\end{equation}

 Instead of the eight 
parameters of equation (\ref{eq:Non_physical_basis}), a more convenient choice 
would be:
\begin{equation}
\label{eq:Physical_basis}
M_{h^0}\,, M_{H^0}\,, M_{A^0}\,, M_{H^\pm}\,, \alpha\,, \tan\beta\,,  m^2_{12}, 
\end{equation}
where $\alpha$ is the mixing angle that rotate the non-physical states
$\phi_1^0$ and $\phi_2^0$ to the physical ones $h^0$ and $H^0$.

For the Yukawa Lagrangian, if we proceed like in the SM and ask that both Higgs doublets couple to all SM fermions, we end up with neutral Higgs couplings to fermions that are flavor violating, i.e., the aforementioned FCNCs. The existence of such interactions would induce large contribution to low energy observables such as $B$, $D$, and $K$ meson oscillations and others which could contradict precise experimental measurements. To evade this problem, one can advocate so-called Natural Flavour Conservation by imposing the aforementioned discrete symmetry,  
$Z_2$  \cite{Glashow:1976nt}, which forbids them. As intimated, to achieve a 2HDM dynamics at once compliant with EWSB and FCNC limits yet offering a (pseudo)scalar mass spectrum which is experimentally interesting, we allow for a soft $Z_2$ breaking (i.e., a small $m_{12}^2$ term).

There exist several ways of coupling the two Higgs doublets to the SM fermions. In the 2HDM Type-I, $\Phi_2$ couple to all fermions just like in the SM while the other $\Phi_1$ does not\footnote{See \cite{Branco:2011iw} for a review of the 2HDM covering other Yukawa types.}. The discrete symmetry behind this assignment is $\Phi_1 \to -\Phi_1$.  The Yukawa Lagrangian of the Type-I model is 
\begin{equation}
\label{eq:Yukawa_Lagrangian}
\mathcal{L}_{\rm Yukawa} = Y_{d}{\overline Q}_L\Phi_2d_R^{} 
                         + Y_{u}{\overline Q}_L\tilde{\Phi}_2u_R^{}
                         + Y_{e}{\overline L}_L\Phi_2 e_R^{}+{H.c.},
\end{equation}
where ${\overline Q}_L$ and ${\overline L}_L$ are the left-handed quark and 
lepton doublets, $d_R$, $u_R$ and $e_R$ are the right-handed up-type quark, 
down-type quark and lepton singlets, respectively, 
$Y_{u}$, $Y_{d}$ and $Y_{e}$ 
are the corresponding Yukawa coupling matrices and 
$\tilde{\Phi}_2=i\sigma_2\Phi^*_2$ (where $\sigma_2$ is the second Pauli matrix). After expressing the weak eigenstates of $\Phi_2$ in terms of the 
physical ones, the Yukawa Lagrangian in Eq. (\ref{eq:Yukawa_Lagrangian}) becomes:
\begin{eqnarray}
 - {\mathcal{L}}_{Yukawa} = \sum_{\psi=u,d,l} \left(\frac{m_\psi}{v} \kappa_\psi^h \bar{\psi} \psi h^0 + 
 \frac{m_\psi}{v}\kappa_\psi^H \bar{\psi} \psi H^0 
 - i \frac{m_\psi}{v} \kappa_\psi^A \bar{\psi} \gamma_5 \psi A^0 \right) + \nonumber \\
 \left(\frac{V_{ud}}{\sqrt{2} v} \bar{u} (m_u \kappa_u^A P_L +
 m_d \kappa_d^A P_R) d H^+ + \frac{ m_l \kappa_l^A}{\sqrt{2} v} \bar{\nu}_L l_R H^+ + H.c. \right),
 \label{Yukawa-1}
\end{eqnarray}
where $\kappa_i^S$ are the Yukawa couplings in the 2HDM. 
The values of those couplings in the 2HDM Type-I are given in 
Tab. (\ref{Yukawa-one}),

\begin{table}
 \begin{center}
  \begin{tabular}{|l|l|l|l|l|l|l|l|l|l||}
   \hline 
    $\kappa_u^h$ & $\kappa_d^h$ & $\kappa_l^h$ & $\kappa_u^H$ & $\kappa_d^H$ & $\kappa_l^H$ & $\kappa_u^A$ & $\kappa_d^A$ & $\kappa_l^A$ \\ \hline
   $c_\alpha/s_\beta$ & $c_\alpha/s_\beta$& $c_\alpha/s_\beta$ & $s_\alpha/s_\beta$ & $s_\alpha/s_\beta$ & $s_\alpha/s_\beta$ & $c_\beta/s_\beta$ & 
    $-c_\beta/s_\beta$ & $-c_\beta/s_\beta$ \\ \hline
    \end{tabular}
 \end{center}
 \caption{Yukawa couplings in the 2HDM Type-I.}
 \label{Yukawa-one}
\end{table}

From the kinetic terms of the Higgs doublets, one can derive the interactions between the gauge bosons and a pair of Higgs (pseudo)scalars, such as 
$H^\pm h^0 W^\pm$, $H^\pm A^0 W^\pm$ and $H^\pm H^0 W^\pm$.  
The corresponding couplings are determined by the gauge coupling structure as well as the angles $\alpha$ and $\beta$ \cite{haber}. They are given by the following relations:
\begin{align}
& g_{H^\pm h^0 W^\pm}=\frac{g\cos(\beta-\alpha)}{2}(p_{h^0}-p_{H^\pm})^\mu,\\
& g_{H^\pm H^0 W^\pm}=\frac{g\sin(\beta-\alpha)}{2}(p_{h^0}-p_{H^\pm})^\mu,\\
& g_{H^\pm A^0 W^\pm}=\frac{g}{2}(p_{A^0}-p_{H^\pm})^\mu,
\end{align}
with $p_\mu$ being the incoming momentum for the corresponding particle.\\

If the charged Higgs boson is light, the top quark can decay into either $W^\pm b$ or  $H^\pm b$. The first decay is controlled by a SM gauge coupling
while the second decay depends upon $\beta$ and it can be enhanced for small $\tan\beta$.
Conventionally, a light charged Higgs boson is assumed to  
decay into either $\tau\nu$ or $cs$, with the corresponding couplings given in Tab. I. However, if  there is an additional light neutral Higgs boson $h^0$ or $A^0$, the additional decay channels into $A^0W^{\pm(*)}$ and $h^0W^{\pm(*)}$ would open up, wherein the $W^\pm$ boson can be on- or off-shell depending on the mass differences $M_{H^\pm}-M_{A^0}$ and $M_{H^\pm}-M_{h^0}$, respectively.  
The $H^\pm \to h^0 W^{\pm(*)}$ channel for a light charged Higgs bosn is open only if we demand that  $H^0$ is the observed 125 GeV SM-like Higgs state (which we do here). In this case, $\vert \cos(\beta-\alpha)\vert\sim 1$ is preferred by experiments and thus the $H^\pm h^0 W^\pm$ couplings is large. The $H^\pm A^0W^\pm$ couplings is independent of $\sin(\beta-\alpha)$ and thus always large. Finally, the $H^\pm\to H^0W^{\pm(*)}$ decay channel is greatly suppressed since it is proportional to $|\sin(\beta-\alpha)|^2$ and the mass difference involved 
($M_{H^\pm}-M_{H^0}$) could be very small since  $M_{H^\pm}<m_t$ and $M_{H^0}= 125$ GeV.

To study our 2HDM Type-I,  we perform a systematic numerical scan over its parameter space as illustrated in Tab.~\ref{tab:2HDM_parameter}. 
During the scan,  each sampled point is subjected to a set of theoretical and experimental constraints which are described in the following section.

\begin{table}[H]
    \centering
    \begin{tabular}{|c|c|}
        \hline
        Parameters & Ranges \\
        \hline
        $M_{h^0}$ & [$10$ , $120$] GeV \\
        $M_{A^0}$ & [$10$ , $120$] GeV \\
        $M_{H^\pm}$ & [$80$ , $170$] GeV \\
        $\sin(\beta-\alpha)$ & [$-0.3$ , $-0.05$] \\
        $m_{12}^2$ & [$0$ , $M_{H^0}^2\sin\beta\cos\beta$] GeV$^2$ \\
        $\tan\beta$ & [$1$ , $60$] \\
        \hline
    \end{tabular}
    \caption{2HDM parameter ranges adopted for our scan. (Note that we have taken $M_{H^0}=125$ GeV.)}
 \label{tab:2HDM_parameter}
\end{table}

\section{Theoretical and Experimental Constraints} 
 
The 2HDM parameters (\ref{eq:Physical_basis}) are constrained by a number of 
theoretical considerations such as vacuum stability  \cite{Deshpande:1977rw}, 
perturbativity, perturbative unitarity \cite{Akeroyd:2000wc}  and experimental limits from LEP, Tevatron, LHC as well as $B$ physics experiments. 
We list here the constraints that we have used.

From the theoretical side, we have the following.
\begin{itemize}
\item A necessary condition for the stability of the vacuum comes from requiring that the scalar potential remain bounded from below when the (pseudo)scalar fields become large. This should be fulfilled in any arbitrary direction 
  in the field space.
In the limit of large field values, the scalar potential is dominated by quartic couplings and one can show that the tree-level vacuum stability constraints are  \cite{Deshpande:1977rw} 
         \begin{equation}
             \lambda_1>0,\quad \lambda_2>0,\quad 
             \lambda_3>-\sqrt{\lambda_1\lambda_2},\quad 
             \lambda_3+\lambda_4-|\lambda_5|>-\sqrt{\lambda_1\lambda_2}\,.
         \end{equation}

\item We require that all $\lambda_i$ remain perturbative and satisfy
             $ |\lambda_i| \le 8 \pi \,. $

\item Perturbative unitarity constraints  \cite{Akeroyd:2000wc} are those obtained by requesting that the $S$-wave component of the  various (pseudo)scalar-(pseudo)scalar, (pseudo)scalar-gauge boson and gauge-gauge bosons scatterings remains unitary at high energy. Such a condition implies a set of constraints that have to be fulfilled and are 
given in \cite{Akeroyd:2000wc}.
    
\item We also check the consistency at 95\% Confidence Level (CL) with the experimental measurements of the oblique parameters $S$, $T$ and $U$. 
We compare those to the fit values of \cite{oblique}, i.e., 
$S= 0.05\pm0.11$,  $T=0.09\pm0.13$ and $U=0.01\pm0.11$. 
  
\end{itemize}
Note that unitarity, perturbativity, vacuum stability  as well as $S, T$ and $U$ constraints are enforced through the 2HDMC public code \cite{thdmc}.

From the experimental side,  we considered the following.

\begin{itemize}
\item $B$-physics observables
 are implemented with the  code 
{SuperIso~v4.0}~ \cite{Mahmoudi:2008tp}. Specifically, we have used the following 
measurements:

\begin{enumerate}
\item $\left.{\rm BR}(B \to X_s\gamma)\right|_{E_\gamma>1.6~\mathrm{GeV}}=(3.32 \pm 0.3)\times 10^{-4}$~ \cite{Amhis:2014hma,HFAG:btosg},

\item ${\rm BR}(B_s \to \mu\mu)=(3.1 \pm 1.4)\times 10^{-9}$~ \cite{Amhis:2014hma,HFAG:BsRare},

\item ${\rm BR}(B^+ \to \tau^+ \nu_{\tau}) = (1.06^{+0.38}_{-0.28})\times 10^{-4}$~ \cite{Amhis:2014hma,HFAG:Bplus}.

\end{enumerate}
For all such observables we allow a 2$\sigma$ tolerance from  the above measurements.
\end{itemize}

\begin{itemize}
\item Consistency with the $Z$ width measurement from LEP, $\Gamma_Z=2.4952\pm0.0023$ GeV  \cite{Tanabashi:2018oca}. Specifically, the partial width $\Gamma(Z\to h^0A^0)$ was required to fall within the $2\sigma$ experimental uncertainty of the measurement ($\leq 4.6$ MeV).

\item Consistency of the mass and signal rates of $H^0$ with the LHC data on $H_{\rm obs}$. We require that the relevant quantities, calculated with {HiggsSignals-v2.2.0beta} \cite{Bechtle:2013xfa}, satisfy these measurement at 95\% CL, assuming Gaussian uncertainties.

\item Consistency of all Higgs states with the direct search
    constraints from LEP, Tevatron and LHC at the 95\% CL, which are 
tested using the updated version of {HiggsBounds-5.3.2beta} \cite{Bechtle:2013wla}. 

\end{itemize}

\section{Signatures}

In this section, we present the results of the aforementioned scans, for the purpose of selecting BPs amenable to MC analysis.

\begin{figure}[H]
\begin{center}
        \includegraphics[width=12cm]{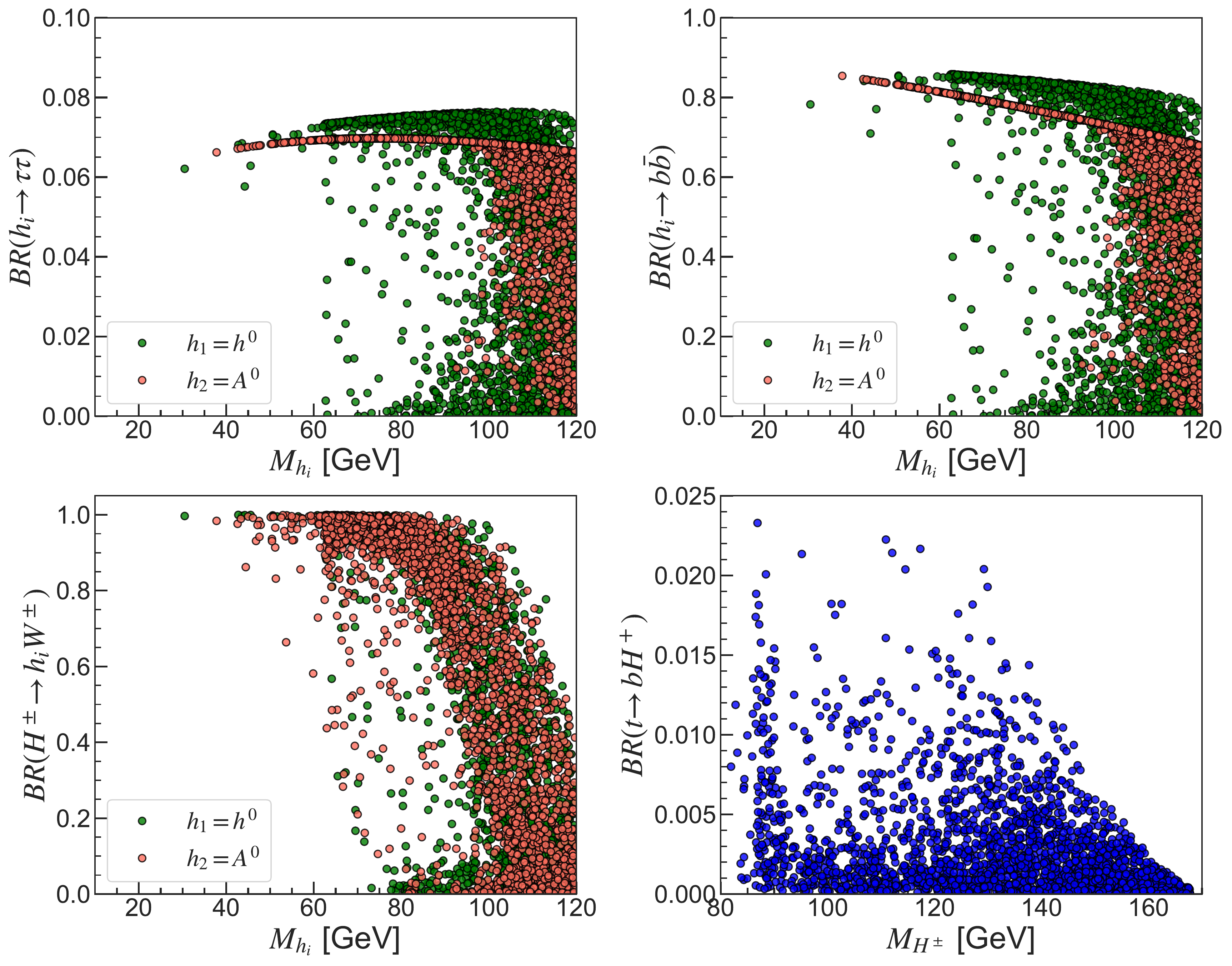}
    \end{center}
\caption{(Top-left) BRs of neutral Higgs ($h^0/A^0\to \tau^+\tau^-$.  (Top-right) ${\rm BR}(h^0/A^0\to b\bar{b}) $.
(Bottom-left) $H^\pm \to h^0/A^0W^{\pm(*)}$. (Bottom-right) ${\rm BR}(t\to bH^+)$.   BRs are given as a function of the relevant Higgs masses.}
\label{fig:fig1}
\end{figure}

We first illustrate the BRs of $h^0$, $A^0$, $H^\pm$ and $t$ into different final states. We start with  Fig.~\ref{fig:fig1}. In the top two frames,
we present BR$(h^0/A^0\to \tau^+\tau^-)$ (left) and  
${\rm BR}(h^0/A^0\to b\bar{b}) $ (right). 
It is clear that the ${\rm BR}(h^0/A^0\to \tau^+\tau^-)$ could reach at best 8\%
 while  ${\rm BR}(h^0/A^0\to b\bar{b})$ can reach up to 80\%. However, the former final state is much cleaner than the latter one in the LHC environment. In this connectoin, upon recalling that the full decay chain $H^\pm\to h^0/A^0 W^{\pm(*)}\to b\bar b W^{\pm(*)}$ is subject to interference effects with $H^\pm\to bt^*\to b\bar b W^{\pm(*)}$ and significant background from $t\bar t$ production and decay \cite{Moretti:2016jkp,Moretti:2016sod,Arhrib:2017veb}, we are induced to elect as tentative signal in our analysis the case of $\tau$ final states. 
Furthermore, in the bottom-left frame of  Fig.\ref{fig:fig1}, we demonstrate that the bosonic 
 decays of the charged Higgs boson, ${\rm BR}(H^\pm \to h^0/A^0W^{\pm(*)})$, could become 
 the dominant ones, even reaching  100\% in some cases (see also  \cite{Arhrib:2016wpw}). Finally,  in the bottom-right frame of Fig.~\ref{fig:fig1}, we present ${\rm BR}(t\to bH^+)$, illustrating the fact that this decay channel of the top (anti)quark is not excessively suppressed with respect to  the SM one $t\to bW^+$. We therefore recommend the $t\to bH^\pm\to b h^0/A^0W^{\pm(*)}\to b\tau^+\tau^-W^{\pm(*)}$ decay chain as the  one to be pursued experimentally, assuming either single or double top (anti)top hadro-production.

\begin{figure}[H]
    \begin{center}
        \includegraphics[width=7cm]{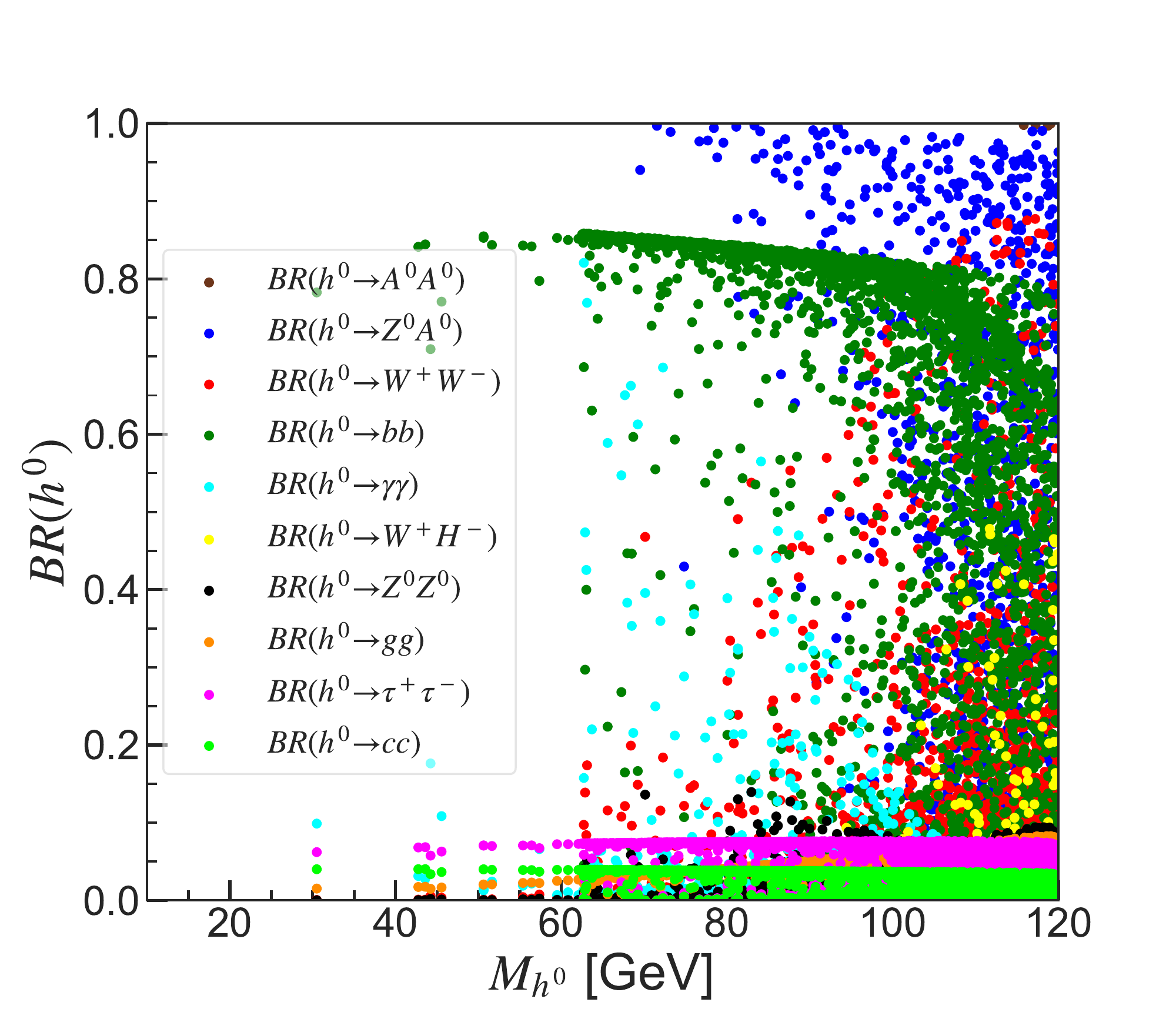}
        \includegraphics[width=7cm]{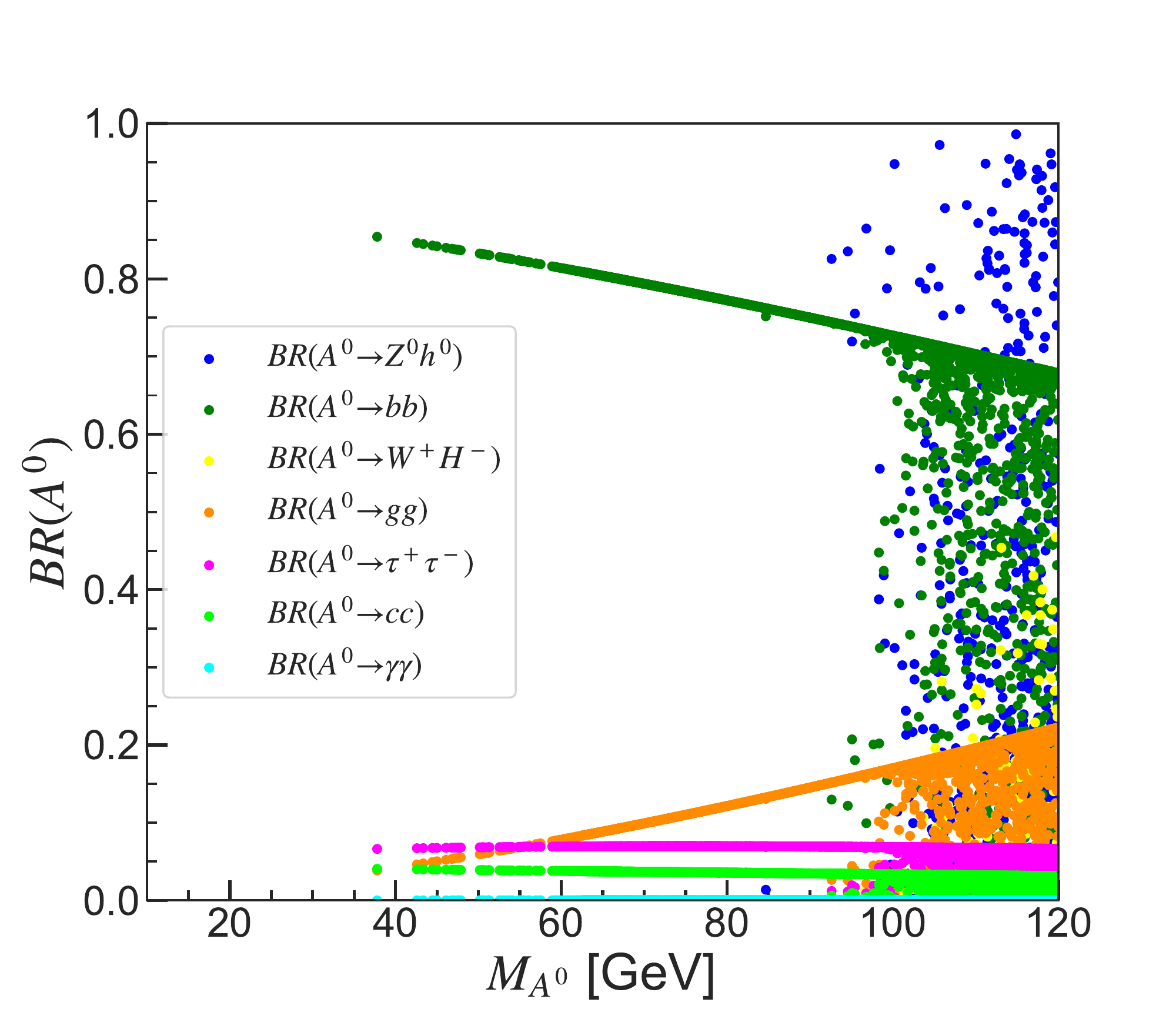}
    \end{center}
    \caption{BRs of $h^0$ (left)  and  $A^0$ (right) as a function of its   mass.}
\label{fig:fig2} 
\end{figure}

A further advantage of the   $\tau^+\tau^-$ decay is that its rate is very independent of the $h^0$ and $A^0$ masses, thus enabling the implementation of a model-independent selection,  
unlike any other decay\footnote{With the exception of the $c\bar c$ one, but this mode has a lower BR and a much larger background in the QCD polluted environment of the LHC, hence of little use here.}, as seen in Fig.~\ref{fig:fig2}. This implies that there is no loss of sensitivity through the $\tau^+\tau^-$ mode in an experimental search, no matter the actual value oh $m_{h^0}$ or $m_{A^0}$. However, other $h^0$ and/or $A^0$ decay modes may well be useful. In fact,  one  can see
from Fig.~\ref{fig:fig2} (left) 
that $h^0$ could become fermiophobic in some cases
making ${\rm BR}(h^0\to \gamma \gamma)$ 
 \cite{Arhrib:2017wmo,Arhrib:2017uon} or 
${\rm BR}(h^0\to ZA^0)$ the dominant decay mode. Further, from  
Fig.~\ref{fig:fig2} (right), it is clear that the suppression of $A^0\to b\bar{b}$ for large $\tan\beta$
allows for a substantial enhancement of $A^0\to Zh^0$ and/or  $A^0\to W^{\pm(*)} H^\mp$. However,  the distributions of these alternative decay rates are rather patchy over the parameter space, so that a model-dependent search would be required to maximize the experimental sensitivity, so we do not consider these in our MC analysis.  

To quantify the size of the light charged Higgs cross-section from the single and double top (anti)quark production followed by the discussed decay chains,
we evaluate the quantities $X^{tj}(h_i,f\bar f)$ and $X^{tt}(h_i,f\bar f)$, respectively, which are defined as follows:
\begin{align}
    X^{tj}(h_i,f\bar f) &= \sigma(pp \to tj) \times {\rm BR}(t \to bH^+) \times {\rm BR}(H^+ \to h_iW^+) \times {\rm BR}(h_i \to f\bar f), \\
    X^{tt}(h_i,f\bar f) &= \sigma(pp \to tt) \times {\rm BR}(t \to bH^+) \times {\rm BR}(t \to bW^+) \times {\rm BR}(H^+ \to h_iW^+) \times {\rm BR}(h_i \to f\bar f).
\end{align}
{{(Notice that the subprocesses contributing to $tj$ production are $bq\to t q'$ and $q\bar q'\to t b$ while those entering $tt$ production are  $q\bar q,gg\to t\bar t$.)}}
\begin{figure}[H]
    \begin{center}
        \includegraphics[width=7cm]{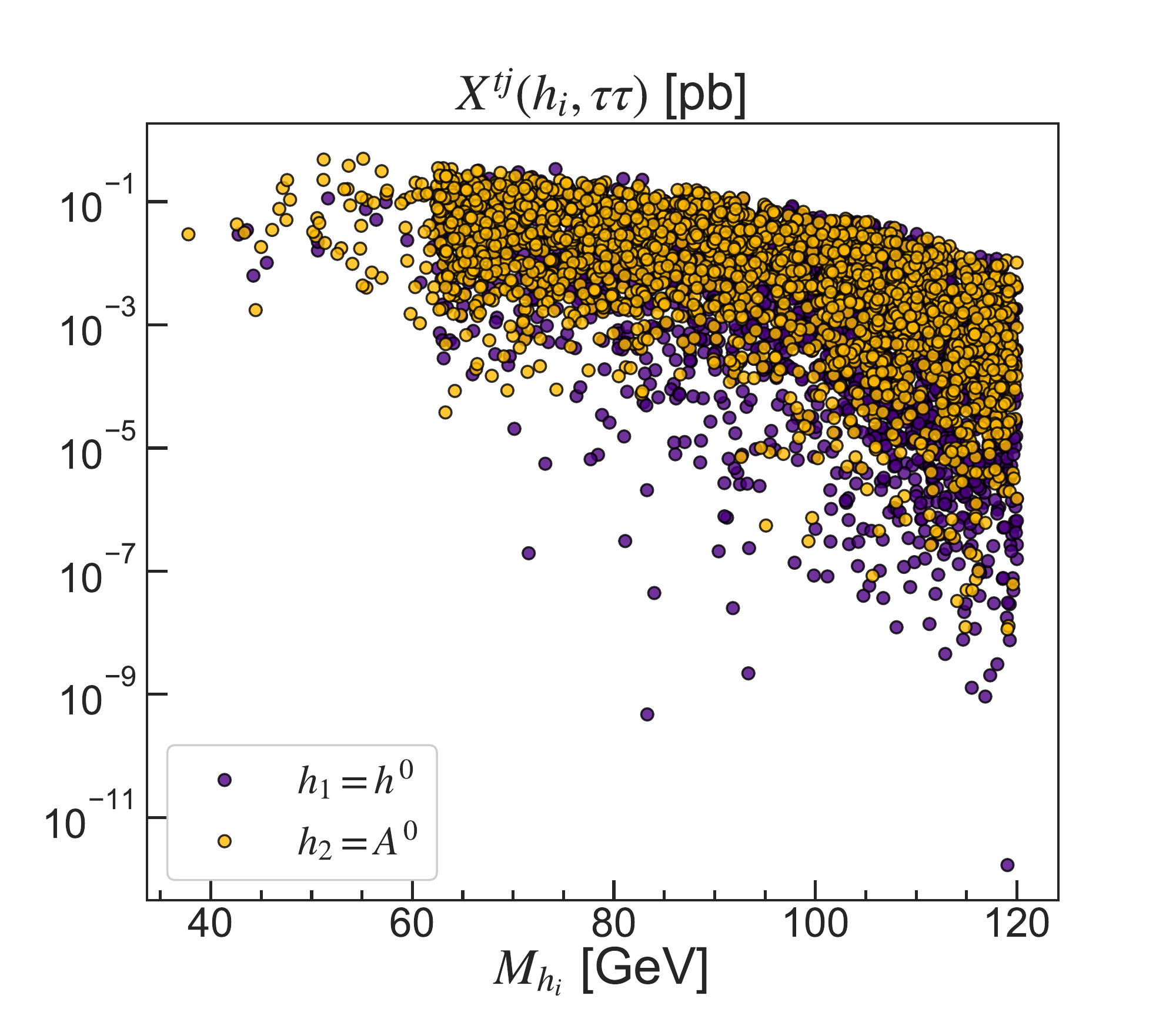}
        \includegraphics[width=7cm]{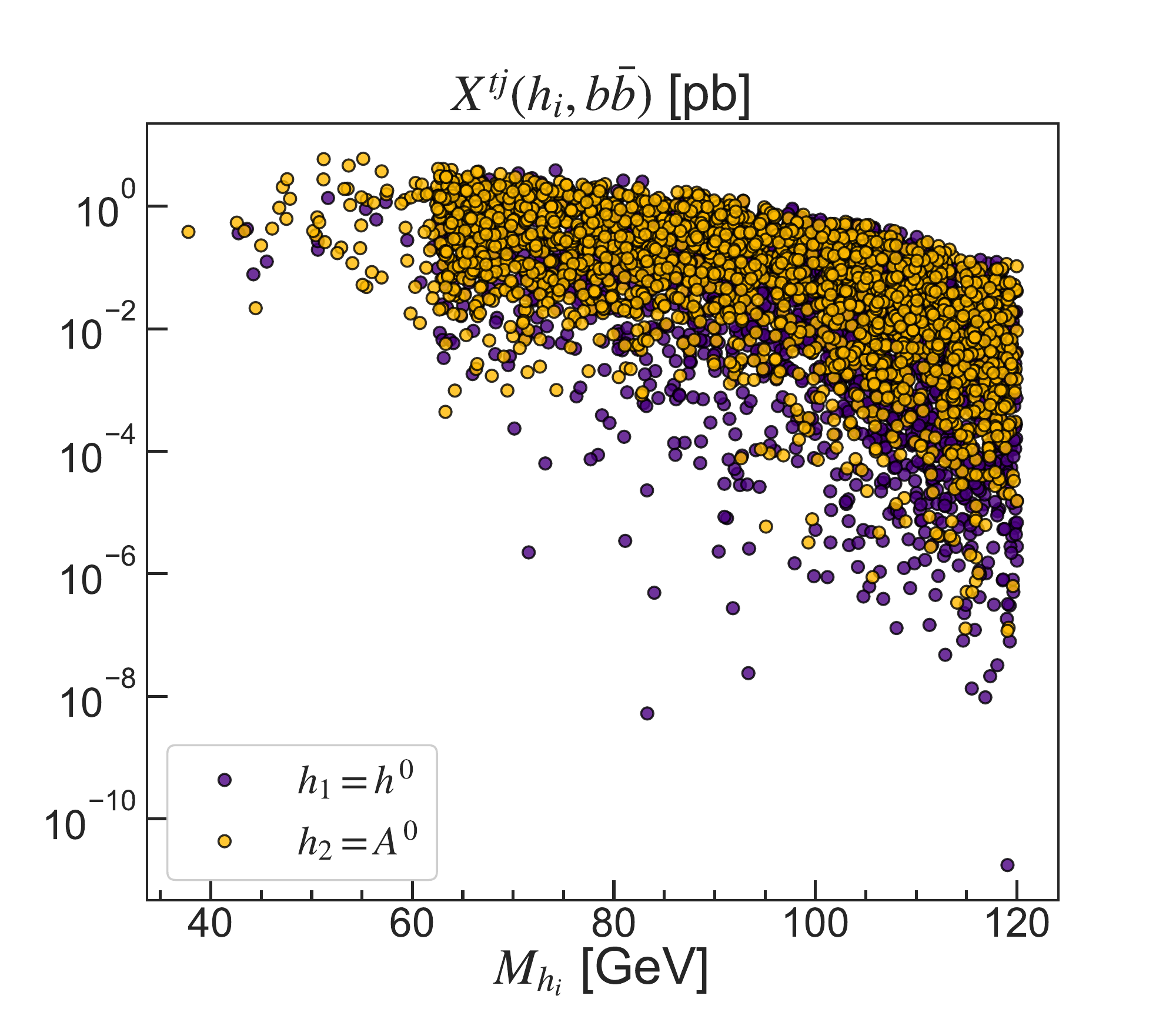}
    \end{center}
    \caption{Cross section for $\sigma(pp\to tj)\times {\rm BR}( t\to H^{\pm}b) \times {\rm BR}(H^\pm \to h_iW^{\pm}) \times {\rm BR}(h_i\to ff) $ as a function of 
    the $h^0$ and $A^0$ mass with $ff=\tau^+\tau^-$ (left) and  $ff=b\bar{b}$ (right).}
    \label{fig:fig4} 
\end{figure}
\begin{figure}[H]
    \begin{center}
        \includegraphics[width=7cm]{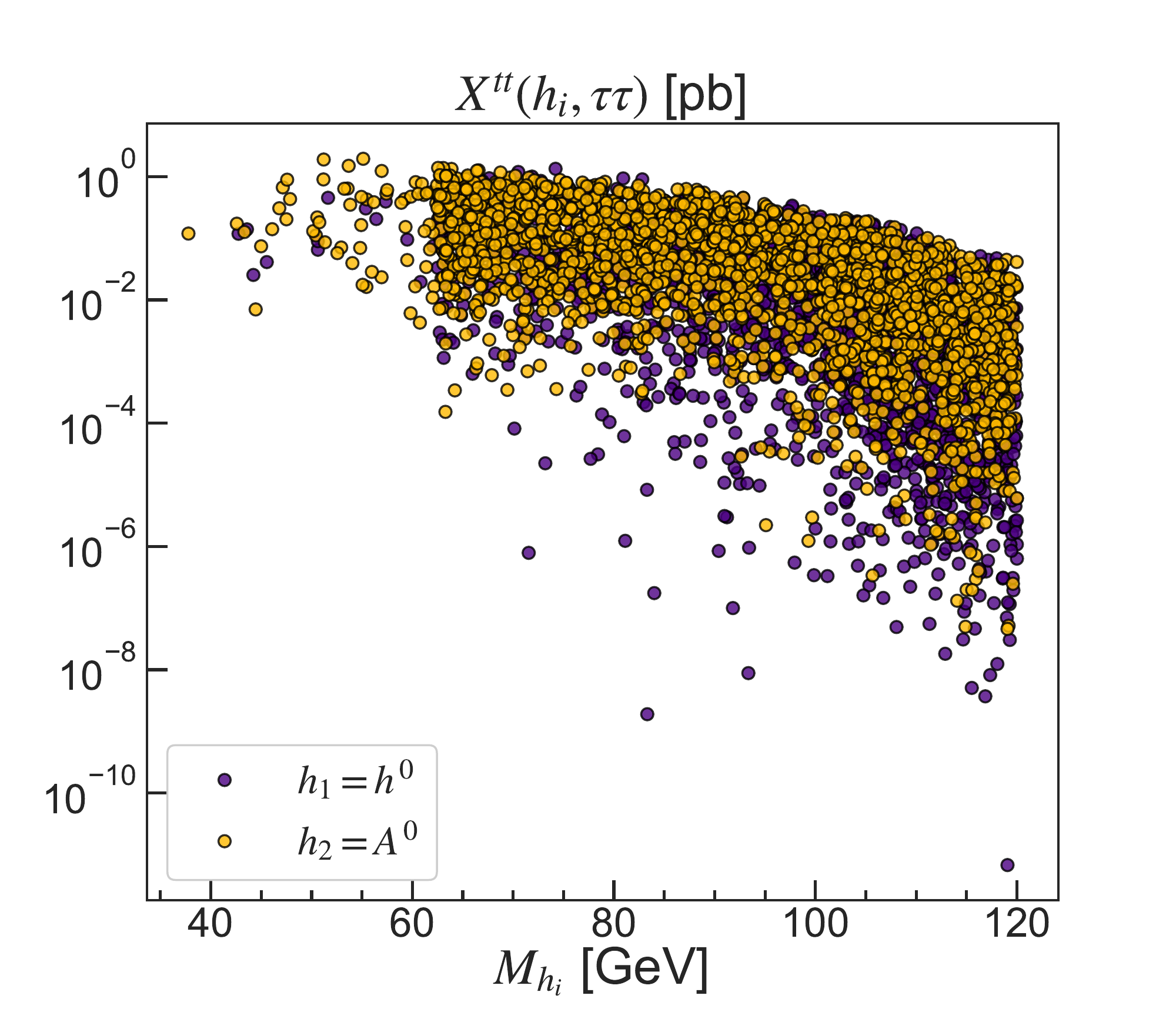}
        \includegraphics[width=7cm]{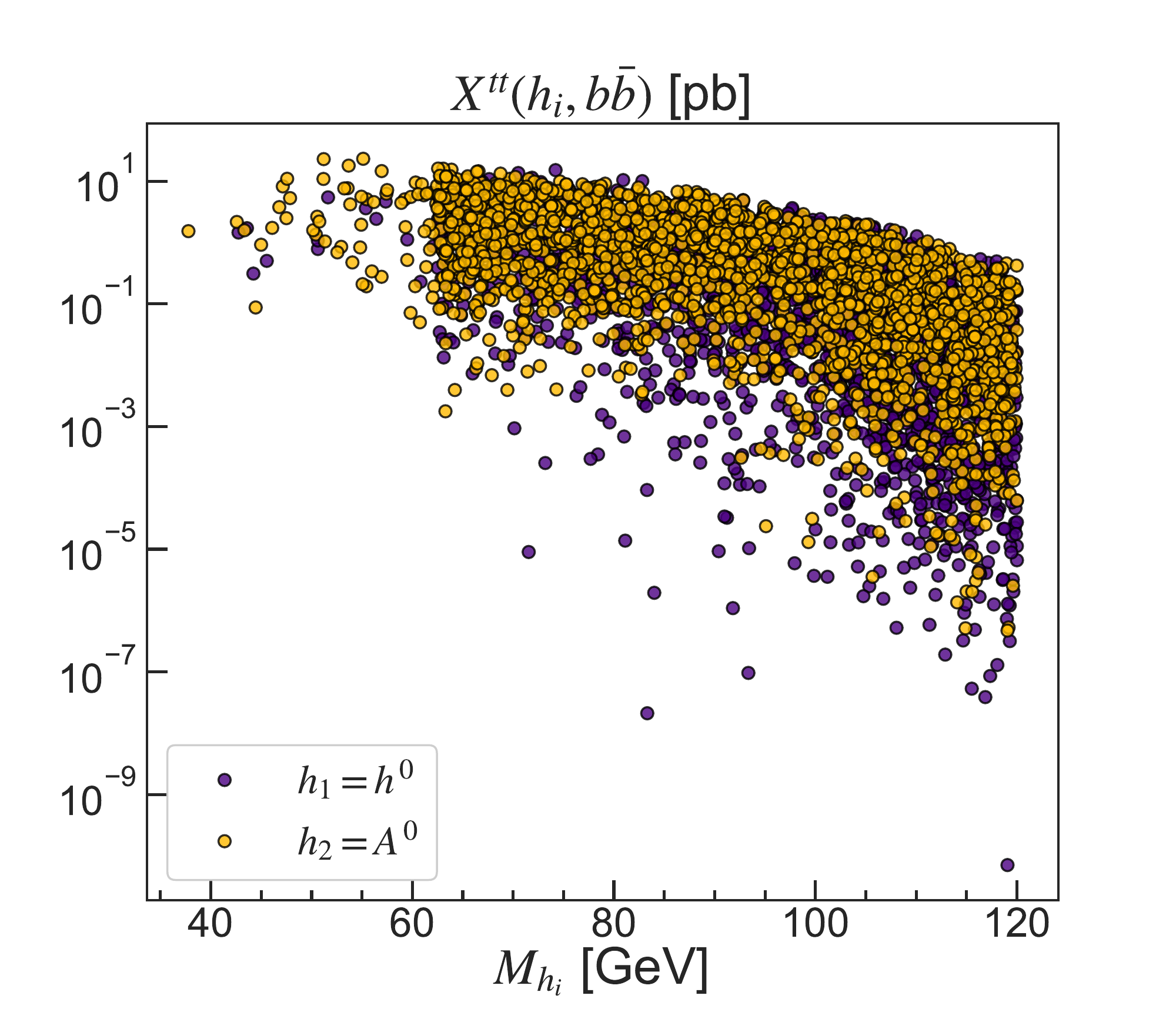}
    \end{center}
    \caption{Cross section $\sigma(pp\to t\bar{t})\times 
    {\rm BR}( t\to H^{\pm}b)\times {\rm BR}( t\to W^{\pm}b)\times {\rm BR}(H^\pm \to h_iW^{\pm}) \times {\rm BR}(h_i\to ff) $ as a function of  the $h^0$ and $A^0$ mass with $f\bar f=\tau^+\tau^-$ (left) and $f\bar f=b\bar{b}$ (right).}
     \label{fig:fig5} 
\end{figure}
In Figs.~\ref{fig:fig4} and \ref{fig:fig5}, we give the numerical results from our scan for both $X^{tj}(h_i,f\bar f)$ and $X^{tt}(h_i,f\bar f)$, respectively, for  
$h^0$ and $A^0$ decaying into $\tau^+\tau^-$ (left frames) or $b\bar{b}$ (right frames).i.e., $ff=\tau^+\tau^-$ and 
$ff=b\bar{b}$ in turn. While the bottom decay rates are dominant over the tau ones, it is clear that the production of charged Higgs bosons from the sum of single and double (anti)top quark followed by either of our bosonic decay modes could reach more than few pb in the $\tau^+\tau^-$ channel. While this value is handsomely large at the inclusive level, it should be recalled that the $W^{\pm(*)}$ boson emerging from the $H^\pm\to h^0/A^0 W^{\pm(*)}$ decay would be searched for in its leptonic transitions, i.e., $ W^{\pm(*)}\to l^\pm\nu$, which are of oder 20\% only, so this calls for using all possible decay patterns of the $\tau^+\tau^-$ pair, i.e., fully hadronic, fully leptonic and semi-leptonic (or semi-hadronic), in order to maximise experimental sensitivity. These three signatures are therefore what we will pursue in our MC study.
 
\subsection{BPs}
\begin{table}[!t]
    \centering
    \scriptsize
    {\renewcommand{\arraystretch}{1}
    {\setlength{\tabcolsep}{0.05cm}
    \begin{tabular}{|c|c|c|c|c|c|c|}
        \hline
        Parameters & BP1 & BP2 & BP3 & BP4 & BP5 & BP6 \\
        \hline
        \hline
        \multicolumn{7}{|c|}{2HDM inputs. The Higgs masses are in GeV.} \\
        \hline
        $M_{h^0}$ & $80.772$ & $78.284$ & $85.003$ & $115.16$ & $119.21$ & $111.13$ \\
        $M_{H^0}$ & $125$ & $125$ & $125$ & $125$ & $125$ & $125$ \\
        $M_{A^0}$ & $116.34$ & $116.14$ & $109.45$ & $64.547$ & $72.896$ & $62.679$ \\
        $M_{H^\pm}$ & $124.29$ & $112.8$ & $132.6$ & $117.23$ & $132.05$ & $98.4$ \\
        $\sin(\beta-\alpha)$ & $-0.073282$ & $-0.065491$ & $-0.077173$ & $-0.13305$ & $-0.086473$ & $-0.057954$ \\
        $\lambda_6$ & $0$ & $0$ & $0$ & $0$ & $0$ & $0$ \\
        $\lambda_7$ & $0$ & $0$ & $0$ & $0$ & $0$ & $0$ \\
        $m_{12}^2$ & $1045.1$ & $676.32$ & $1244.7$ & $2743.6$ & $1979.6$ & $2120.2$ \\
        $\tan\beta$ & $5.5795$ & $6.543$ & $5.1575$ & $4.233$ & $3.5324$ & $5.6988$ \\
        \hline
        \hline
        \multicolumn{7}{|c|}{Total decay width  in GeV} \\
        \hline
        $\Gamma(h^0)$ & $2.465 \times 10^{-5}$ & $1.6301 \times 10^{-5}$ & $3.1343 \times 10^{-5}$ & $0.0001354$ & $0.00019498$ & $0.0001218$ \\
        $\Gamma(H^0)$ & $0.0044033$ & $0.0043866$ & $0.0044133$ & $0.0044791$ & $0.0044727$ & $0.0043936$ \\
        $\Gamma(A^0)$ & $0.00013585$ & $0.00011075$ & $0.00012937$ & $0.00010722$ & $0.0001734$ & $5.752 \times 10^{-5}$ \\
        $\Gamma(H^+)$ & $0.00015084$ & $5.1564 \times 10^{-5}$ & $0.0002483$ & $0.00034353$ & $0.00068742$ & $5.5684 \times 10^{-5}$ \\
        $\Gamma(t)$ & $1.3784$ & $1.3789$ & $1.3765$ & $1.3936$ & $1.3894$ & $1.3903$ \\
        \hline
        \hline
        \multicolumn{7}{|c|}{${\rm BR}(h^0 \to XY)$ in \%} \\
        \hline
        ${\rm BR}(h^0 \to \tau^+ \tau^-)$ & $7.5297$ & $7.5114$ & $7.5603$ & $1.7983$ & $4.8386$ & $2.5942$ \\
        ${\rm BR}(h^0 \to g g)$ & $4.1758$ & $3.9511$ & $4.5778$ & $1.8932$ & $5.4377$ & $2.5543$ \\
        ${\rm BR}(h^0 \to Z^0 A^0)$ & $-$ & $-$ & $-$ & $72.548$ & $35.026$ & $65.059$ \\
        ${\rm BR}(h^0 \to W^+ W^-)$ & $0.029144$ & $0.026408$ & $0.042102$ & $3.4713$ & $1.6768$ & $0.42731$ \\
        ${\rm BR}(h^0 \to c c)$ & $3.8246$ & $3.8413$ & $3.7981$ & $0.84822$ & $2.2665$ & $1.2324$ \\
        ${\rm BR}(h^0 \to b b)$ & $84.271$ & $84.535$ & $83.842$ & $18.872$ & $50.46$ & $27.402$ \\
        \hline
        \hline
        \multicolumn{7}{|c|}{${\rm BR}(H^0 \to XY)$ in \%} \\
        \hline
        ${\rm BR}(H^0 \to \tau^+ \tau^-)$ & $5.9984$ & $5.9905$ & $6.0031$ & $6.0389$ & $6.0262$ & $5.9884$ \\
        ${\rm BR}(H^0 \to g g)$ & $7.3793$ & $7.3696$ & $7.3851$ & $7.4292$ & $7.4135$ & $7.367$ \\
        ${\rm BR}(H^0 \to Z^0 Z^0)$ & $2.3825$ & $2.3942$ & $2.3757$ & $2.3132$ & $2.3406$ & $2.3926$ \\
        ${\rm BR}(H^0 \to W^+ W^-)$ & $19.073$ & $19.166$ & $19.019$ & $18.518$ & $18.737$ & $19.154$ \\
        ${\rm BR}(H^0 \to c c)$ & $2.7834$ & $2.7797$ & $2.7855$ & $2.8022$ & $2.7963$ & $2.7787$ \\
        ${\rm BR}(H^0 \to b b)$ & $62.016$ & $61.934$ & $62.065$ & $62.435$ & $62.304$ & $61.913$ \\
        \hline
        \hline
        \multicolumn{7}{|c|}{${\rm BR}(A^0 \to XY)$ in \%} \\
        \hline
        ${\rm BR}(A^0 \to \tau^+ \tau^-)$ & $5.6976$ & $5.0731$ & $6.5871$ & $6.9509$ & $6.973$ & $6.9414$ \\
        ${\rm BR}(A^0 \to g g)$ & $18.133$ & $16.1$ & $18.93$ & $8.7517$ & $10.538$ & $8.3672$ \\
        ${\rm BR}(A^0 \to Z^0 h^0)$ & $14.745$ & $24.117$ & $2.6261$ & $-$ & $-$ & $-$ \\
        ${\rm BR}(A^0 \to c c)$ & $2.7053$ & $2.4096$ & $3.1664$ & $3.742$ & $3.653$ & $3.762$ \\
        ${\rm BR}(A^0 \to b b)$ & $58.646$ & $52.236$ & $68.611$ & $80.491$ & $78.769$ & $80.866$ \\
        \hline
        \hline
        \multicolumn{7}{|c|}{${\rm BR}(H^+ \to XY)$ in \%} \\
        \hline
        ${\rm BR}(H^+ \to \tau^+ \nu_{\tau})$ & $5.4821$ & $10.583$ & $4.1585$ & $3.9444$ & $3.189$ & $11.267$ \\
        ${\rm BR}(H^+ \to W^+ h^0)$ & $89.931$ & $83.481$ & $87.672$ & $-$ & $0.058026$ & $-$ \\
        ${\rm BR}(H^+ \to W^+ A^0)$ & $0.025258$ & $-$ & $2.7961$ & $93.568$ & $92.76$ & $83.224$ \\
        ${\rm BR}(H^+ \to s c)$ & $2.3214$ & $4.5651$ & $1.7396$ & $1.6889$ & $1.335$ & $4.9918$ \\
        ${\rm BR}(H^+ \to b t)$ & $1.9984$ & $0.89684$ & $3.4528$ & $0.623$ & $2.5191$ & $-$ \\
        \hline
        \hline
        \multicolumn{7}{|c|}{${\rm BR}(t \to XY)$ in \%} \\
        \hline
        ${\rm BR}(t \to b W^+)$ & $98.897$ & $98.865$ & $99.036$ & $97.82$ & $98.115$ & $98.052$ \\
        ${\rm BR}(t \to b H^+)$ & $0.93164$ & $0.96372$ & $0.79287$ & $2.0086$ & $1.7136$ & $1.7766$ \\
        \hline
        \hline
        \multicolumn{7}{|c|}{SuperIso} \\
        \hline
        $\rm{BR}(B \to X_s \gamma)$ & $0.00031283$ & $0.00031418$ & $0.00031207$ & $0.00030838$ & $0.00030483$ & $0.00031247$ \\
        $\rm{BR}(B_s \to \mu^+ \mu^-)$ & $3.2581 \times 10^{-9}$ & $3.2422 \times 10^{-9}$ & $3.2669 \times 10^{-9}$ & $3.3121 \times 10^{-9}$ & $3.3546 \times 10^{-9}$ & $3.2634 \times 10^{-9}$ \\
        $\rm{BR}(B_u \to \tau \nu_\tau)$ & $8.1856 \times 10^{-5}$ & $8.1857 \times 10^{-5}$ & $8.1855 \times 10^{-5}$ & $8.1847 \times 10^{-5}$ & $8.1844 \times 10^{-5}$ & $8.1851 \times 10^{-5}$ \\
        \hline 
        \multicolumn{7}{|c|}{$\sigma\times\rm{BR}$ in pb} \\
        \hline
        $X^{tj}(h^0,\tau^+\tau^-)$ &  $0.156455$ & $0.149869$ & $0.130332$ & $-$ & $0.000119314$ & $-$ \\
        \hline
        $X^{tt}(h^0,\tau^+\tau^-)$ &  $0.62391$ & $0.597452$ & $0.520465$ & $-$ & $0.000472038$ & $-$  \\
        \hline
        $X^{tj}(h^0,b b)$ &  $1.75101$ & $1.68666$ & $1.44536$ & $-$ & $0.00124427$ & $-$ \\
        \hline
        $X^{tt}(h^0,b b)$ &  $6.98264$ & $6.72388$ & $5.77187$ & $-$ & $0.00492267$ & $-$  \\
        \hline
        $X^{tj}(A^0,\tau^+\tau^-)$ &  $3.32498 \times 10^{-5}$ & $-$ & $0.00362156$ & $0.323982$ & $0.274873$ & $0.254534$  \\
        \hline
        $X^{tt}(A^0,\tau^+\tau^-)$ &  $0.000132593$ & $-$ & $0.0144623$ & $1.2779$ & $1.08747$ & $1.00636$ \\
        \hline
        $X^{tj}(A^0,b b)$ &  $0.000342243$ & $-$ & $0.037722$ & $3.75169$ & $3.10504$ & $2.96525$  \\
        \hline
        $X^{tt}(A^0,b b)$ &  $0.00136479$ & $-$ & $0.150638$ & $14.798$ & $12.2843$ & $11.7237$ \\
        \hline
    \end{tabular}}}
    \caption{The full description of our BPs we take $m_t = 173.5$ GeV.}
    \label{tab:benchmarks_points}
\end{table}
In order to perform our MC  simulation, out of our scan, we select six BPs. Detailed information about the latter, including mass spectra and decay BRs, is  presented  in Tab.~\ref{tab:benchmarks_points}. There are several salient features of these BPs that is worth dwelling upon.
\begin{itemize}
\item Both the extra neutral Higgs bosons,  $A^0$ and $h^0$, are lighter than the  discovered Higgs boson, $H^0$.
\item The charged Higgs boson has a mass smaller than the $t$ mass, but the sum of its mass with the mass of an extra neutral Higgs boson (either $A^0$ or $h^0$) is larger than the top quark mass, except for BP6.
\item The rates for ${\rm BR}(t \to b H^\pm) \times  {\rm BR}(H^\pm \to \tau^\pm \nu_{\tau})$ are less than $12 \times 10^{-4}$, i.e.,  comfortably below the current LHC bound.
\item The ${\rm BR}(H^\pm \to A^0 W^{\pm(*)})$ for the first three BPs is equal to $89.9\%$, $83.5\%$ and $87.7\%$, respectively. while the ${\rm BR}(H^\pm \to h^0 W^{\pm(*)})$ for the second three BPs is equal to $93.6\%$, $92.8\%$ and $83.2\%$, respectively. Thus, they are very close to their best possible values (recall Fig.~\ref{fig:fig1}), in turn implying that these are amongst the  2HDM Type-I parameter points most accessible to a future LHC analysis.  
\end{itemize}

We will now proceed to a MC version of the latter.

\section{Collider Analysis}
In this part, we perform a detector level MC simulation to establish the LHC sensitivity to the BPs given in Tab. \ref{tab:benchmarks_points}. Due to the kinematics of these BPs, we will focus on the discovery of charged Higgs boson via the decay $H^{\pm} \to h^{0}l^{\pm}\nu,~l=(e,\mu)$ or $H^{\pm} \to A^{0}l^{\pm}\nu$, via an off-shell $W^{\pm}$ boson. To obtain a meaningful significance in our analysis, we assume the integrated luminosity of 100 $fb^{-1}$ and the collision energy of $\sqrt{s}=14$ TeV at the LHC. 

As mentioned, we will focus on two production channels of the charged Higgs boson, the first is $pp\to tj \to b H^\pm j$ (single top (anti)quark production) and the second is $pp \to t {\bar t} \to bH^\pm t$ (top quark pair production). 

After its production, the charged Higgs boson will decay into $h^0 W^{\pm *}$ or $A^0 W^{\pm *}$, wherein the neutral Higgs state will further decay to $\tautau$. The $\tautau$ decay mode can be categorized into three cases, in term of the final states it produces, i.e., fully hadronically, fully leptonically and semi-leptonically (or semi-hadronically), which we identify as 
$\tau_{\rm had}\tau_{\rm had}$, $\tau_{\rm had}\tau_{\rm lep}$, $\tau_{\rm lep}\tau_{\rm lep}$, which will be described in more detail below for each of the two production channels. Further, because in the following analysis we do not use angular correlations of the decay products of the charged Higgs boson, our selection method can be used for both for $H^{\pm} \to h^{0} l^{\pm}\nu$ and $H^{\pm} \to A^{0}l^{\pm}\nu$ at the same time, so we will treat these two cases in the same manner. The six BPs used for the signals are provided in Tab. \ref{tab:benchmarks_points}.

Now we present our MC  analysis at a detector level, for both signal and background, which has the following features.
\begin{itemize}
\item  We use MadGraph5\_aMC@NLO  v2.6.5 \cite{Alwall:2014hca} to compute the cross-sections and generate both signal and background events at Parton level. At this level, we adopt the following kinematic cuts to improve the efficiency of the MC event generation
\begin{equation}
|\eta(l,j)|<2.5, \quad p_T(l,j)>20 ~\text{GeV}, \quad \Delta R(l,j)>0.5,
\end{equation} 
where $j$ refers here to a Parton.  Further, after generating the signals at the Leading Order (LO), we reweight each corresponding event by using the Next-to-LO (NLO) cross-sections given in Tab.~\ref{tab:benchmarks_points}. 

For the signal processes at the Parton level, the full decay chains are specified. For example, for single top production, we generate the matrix elements of the processes $pp\to t j \to H^\pm b j \to h^0(A^{0}) W^\pm b j \to \tau \tau W^\pm b j$ while, for top pair production, we generate the matrix elements of the processes $pp\to t {\bar t} \to t H^\pm b j \to t h^{0}(A^0) W^\pm b j \to t \tau \tau W^\pm b j$. 

\item After generating both signal and background events at the Parton level, we pass these events to Pythia v6.4  \cite{Sjostrand:2006za} to simulate initial and final state radiation (i.e., the QED and QCD emission), Parton shower,  hadronization, and heavy flavor decays for each of the events. 

\item At the detector level, we use  Delphes v3.4.2  \cite{deFavereau:2013fsa} to simulate the detector effects and the pseudo-dataset generated by it to perform our analysis. To cluster final particles into jets in each event, we adopt the anti-$k_t$ jet algorithm \cite{Cacciari:2008gp} with jet parameter $\Delta R=0.4$ in the FastJet package \cite{Cacciari:2011ma}. 
\end{itemize}

From the final objects reconstructed at the detector level, i.e., $\tau$ jets,  $\tau$ leptons, jets (including $b$-jets), leptons and missing transverse momentum, we perform the signal and background discriminant analysis below. 
%
%
We will consider the signal for the case of single (anti)top quark production and then for the case of double (anti)top quark production.

\subsection{Charged Higgs boson production from $pp \to tj$ processes }\label{tjproduction}

In this subsection, we will study charged Higgs boson production from single top production. Charged Higgs bosons can be produced from the top quark decay, the signal process is thus given as $pp\to tj \to bH^{\pm}j \to bh^{0}(A^{0})l^{\pm}\nu j \to b\tau^+\tau^- l^{\pm}\nu j$. The BR of a top quark decaying to a charged Higgs boson can reach $1-2\%$ or so, as shown by the BPs given in Tab.~\ref{tab:benchmarks_points}.
The main background processes include: 1)  $pp\to W^\pm \tau^+\tau^-$ production, which is generated up to two additional jets;  2) $pp \to t\bar t$ production with fully leptonic, semi-leptonic (or semi-hadronic) and fully hadronic decay modes, which are all generated up to one additional jet; 3) $t\bar t l'l'$ production, where $l'=e,\mu,\tau$; 4) $tj\tau^+\tau^-$ production.  

As intimated, both signal and background events are categorized into three cases in terms of the final state emerging from the $\tautau$ decay.
\begin{itemize}
\item Case A: both $\tau$'s under hadronic decays. Thus, the final state includes two tagged $\tau$ jets and one lepton, $l=e,\mu$, which is from the off-shell $W^\pm$ decay, plus one or two extra jets.

\item Case B: one $\tau$ undergoes hadronic decay, the other $\tau$ undergoes leptonic decay plus there is an extra lepton from the off-shell $W^\pm$ decay, which is expected to be softer than the one stemming from the leptonic $\tau$ decay. Thus, the final state here is made up of two leptons, one tagged $\tau$ jet and one or two extra jets. 

\item Case C: both $\tau$'s undergo leptonic decays so that here the final state is made up of three leptons and one (two) extra jet(s).
\end{itemize}

Below we list some key kinematic features which are useful to distinguish between signal and background events.
\begin{itemize}

\item \textbf{The reconstructed $h^0$ and $A^0$ bosons}

In a signal event, we can combine two $\tau$ jets (Case A), a $\tau$ jet and the hardest lepton (Case B) or the two hardest leptons (Case C) to look for the mass resonance of a $h^0$ (for BPs 1, 2, and 3) or $A^0$ (for BPs 4, 5, and 6).   In Case A, the reconstructed mass should peak near the input mass of  $h^0(A^0)$. In Cases B and C, due to the large fraction of energy in the event taken away by neutrinos from the $\tau$ leptonic decay(s), the reconstructed mass usually peaks at a value smaller than the actual mass of the $h^0(A^0)$ state. 
\begin{figure}[ht]
  \begin{minipage}{0.30\textwidth}
    \begin{center}    
      (a) Case A    
 \includegraphics[height=4cm]{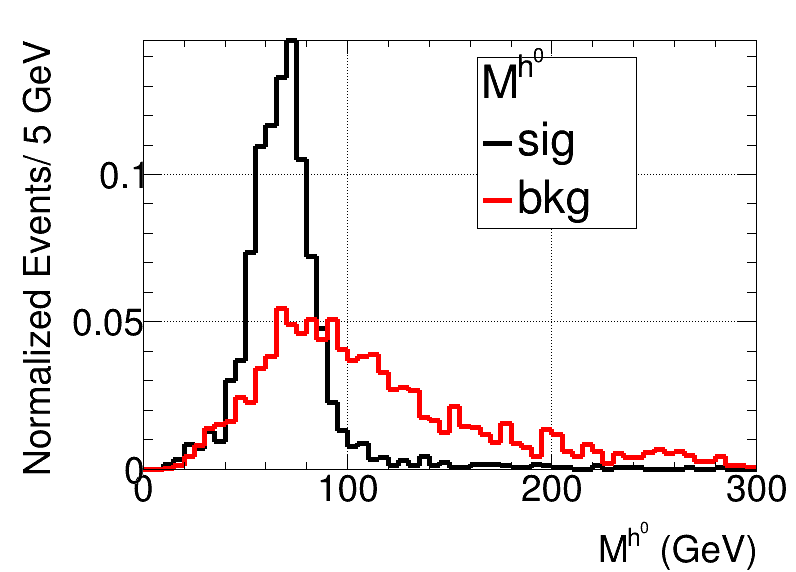} 
  \end{center}
 \end{minipage} 
   \begin{minipage}{0.30\textwidth}
    \begin{center}        
      (b) Case B
 \includegraphics[height=4cm]{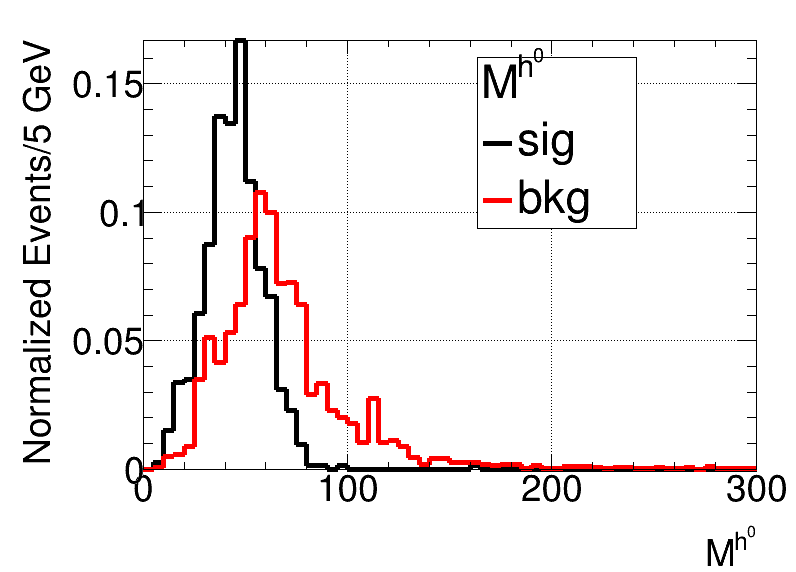} 
  \end{center}
 \end{minipage}
 \begin{minipage}{0.30\textwidth}
    \begin{center}        
      (c) Case C
 \includegraphics[height=4cm]{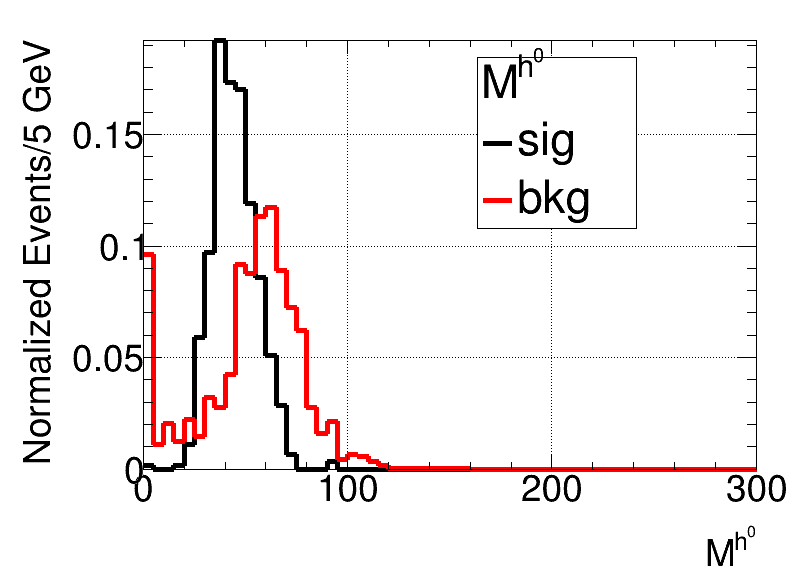} 
  \end{center} 
  \end{minipage}
  \caption{The reconstructed  neutral Higgs boson  mass, $M^{h^0}$,  is shown for Case A (a), Case B (b)  and Case C (c). The signal events are generated by using  BP1 in Tab. \ref{tab:benchmarks_points}. }\label{f_h_mass}
\end{figure} 
To demonstrate this feature, in Fig. \ref{f_h_mass}, we show the distribution of the invariant mass of the two $\tau$ jets for both  signal  and background events, denoted by $M^{h^0}$, where the former  are generated by using BP1 in Tab. \ref{tab:benchmarks_points}, as representative of the typical signal kinematics. 


\item \textbf{The reconstructed $H^{\pm}$ boson}

Since a charged Higgs boson can only transition into the final state $h^{0} l\nu $ or $A^{0} l\nu $ through an off-shell $W^\pm$ decay, we can  attempt reconstructing its mass using the momentum of the lepton, the Missing $E_T$ (MET), where $E_T$ is the missing transverse energy (or momentum), and the reconstructed $h^{0}(A^{0})$. 
\begin{figure}[ht]
  \begin{minipage}{0.30\textwidth}
    \begin{center}    
      (a) Case A    
 \includegraphics[height=4cm]{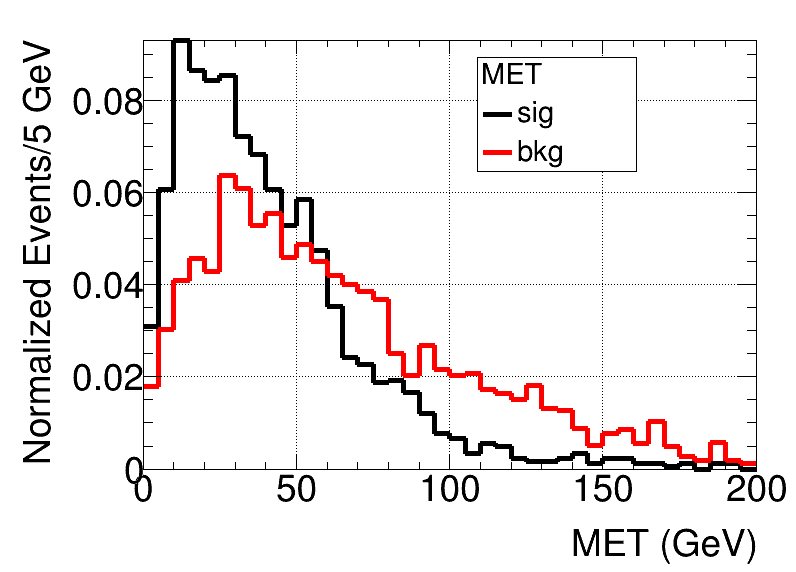} 
  \end{center}
 \end{minipage} 
   \begin{minipage}{0.30\textwidth}
    \begin{center}        
      (b) Case B
 \includegraphics[height=4cm]{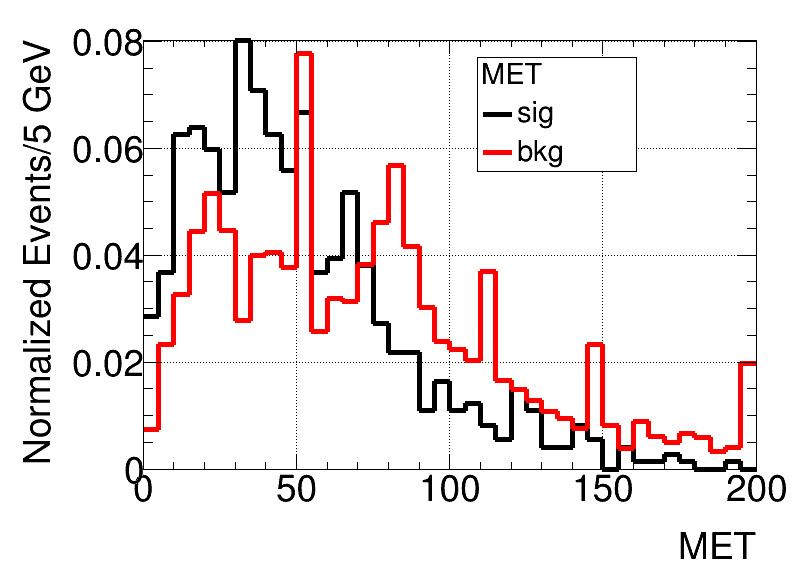} 
  \end{center}
 \end{minipage}
 \begin{minipage}{0.30\textwidth}
    \begin{center}        
      (c) Case C
 \includegraphics[height=4cm]{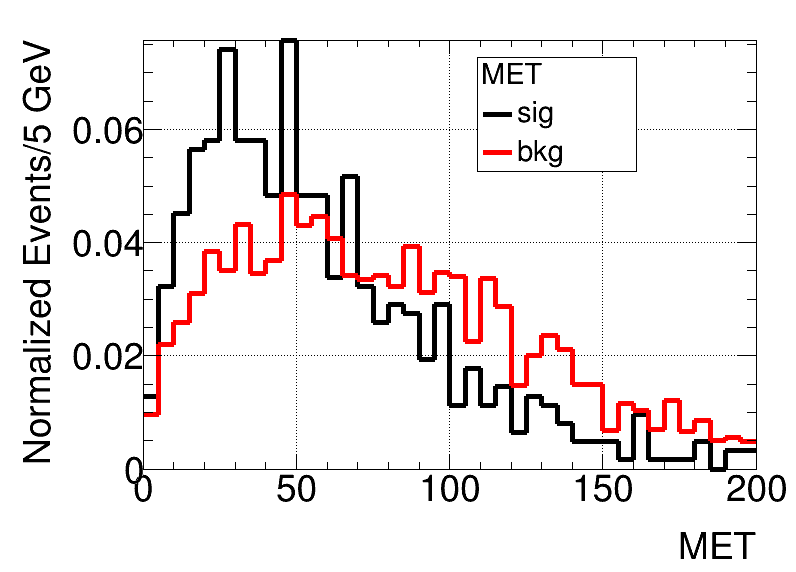} 
  \end{center} 
  \end{minipage}
  \caption{The missing transverse energy, MET,  is shown for Case A (a), Case B (b) and Case C (c). The signal events are generated by using BP1 in Tab. \ref{tab:benchmarks_points}.}\label{f_MET}
\end{figure} 
Notice that the MET  of each event comes from neutrinos from $W^\pm$ decays, including via leptonic $\tau$  decay. In  Fig. \ref{f_MET}, we show its distribution for the usual three cases. In Case A, the MET mainly comes from the $W^\pm$ boson off-shell decay, therefore its peak is at relatively small values. In Cases B and C, leptonic $\tau$ decays also contribute to the MET, so that its peak value  increases.

\begin{figure}[ht]
  \begin{minipage}{0.30\textwidth}
    \begin{center}    
      (a) Case A    
 \includegraphics[height=4cm]{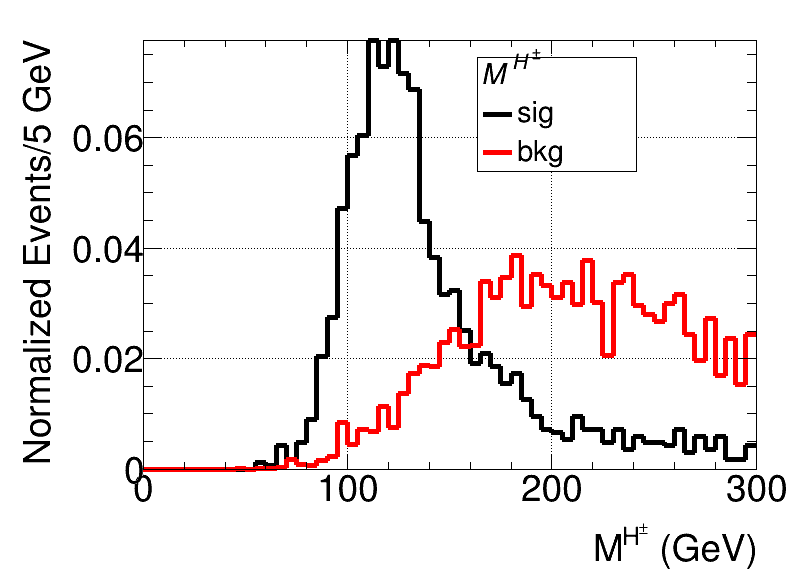} 
  \end{center}
 \end{minipage} 
   \begin{minipage}{0.30\textwidth}
    \begin{center}        
      (b) Case B
 \includegraphics[height=4cm]{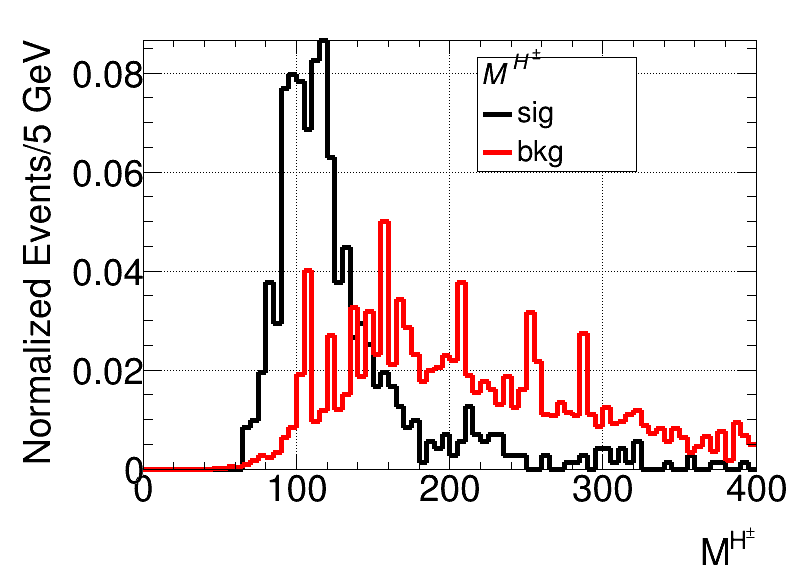} 
  \end{center}
 \end{minipage}
 \begin{minipage}{0.30\textwidth}
    \begin{center}        
      (c) Case C
 \includegraphics[height=4cm]{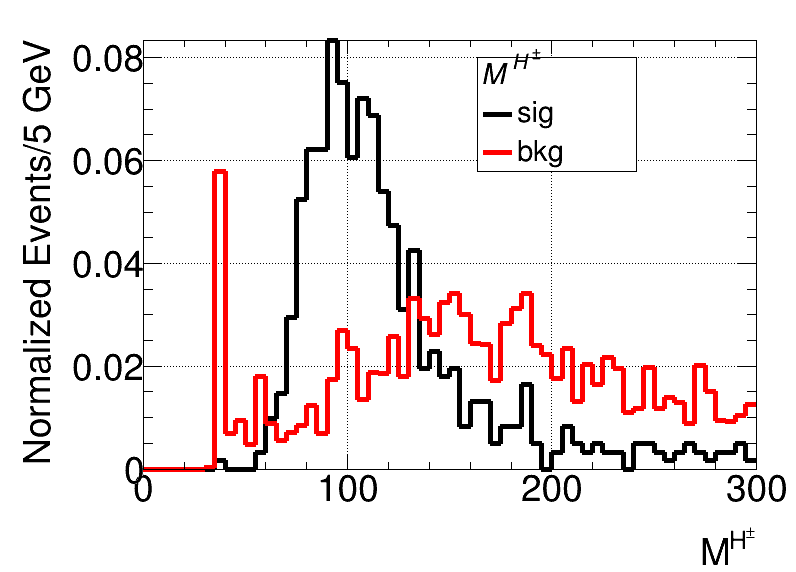} 
  \end{center}
 \end{minipage}
   \caption{The reconstructed charged Higgs boson mass, $M^{H^\pm}$,  is shown for Case A (a), Case B (b) and Case C (c). The signal events are generated by using  BP1 in Tab. \ref{tab:benchmarks_points}.}\label{f_HC_mass}
\end{figure} 

We also use the MET to reconstruct the {{mass of the `$l +$MET$+h^0(A^0)$' system, which we denote by $M^{H^\pm}$, as it is sensitive to the actual value of $M_{H^\pm}$. Based on Ref. \cite{Aad:2012ux}, we proceed as follows.}} We calculate the $\nu$ four-momentum by enforcing the $H^{\pm}$ mass constraint on the $l \nu h^{0}(A^0) $ system and then in turn reconstruct the $H^{\pm}$ four-momentum. The distribution of the ensuing charged Higgs boson mass is shown in Fig. \ref{f_HC_mass}. In all the three cases, it is noticed that the mass peak of the charged Higgs boson can be successfully reconstructed, which can then serve as an efficient discriminant to distinguish between signal and background events. 

\item \textbf{The reconstructed $t$ quark}

\begin{figure}[ht]
  \begin{minipage}{0.30\textwidth}
    \begin{center}    
      (a) Case A    
 \includegraphics[height=4cm]{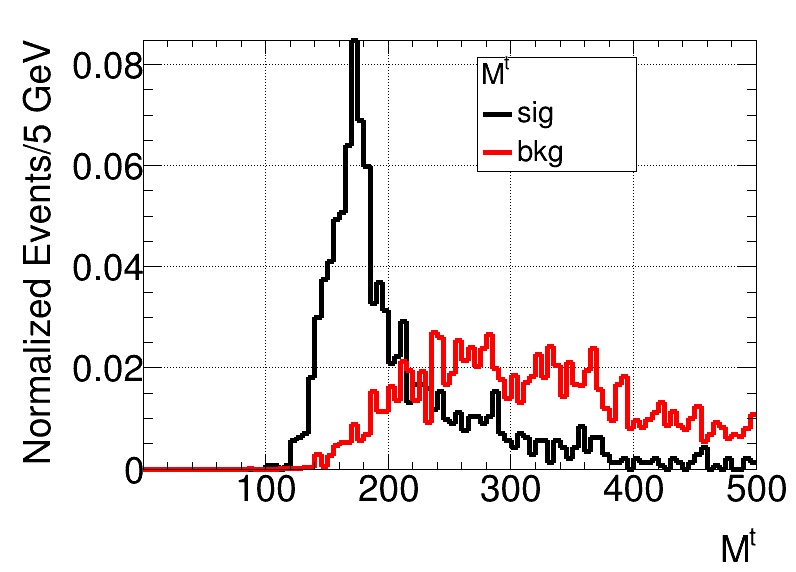} 
  \end{center}
 \end{minipage} 
   \begin{minipage}{0.30\textwidth}
    \begin{center}        
      (b) Case B
 \includegraphics[height=4cm]{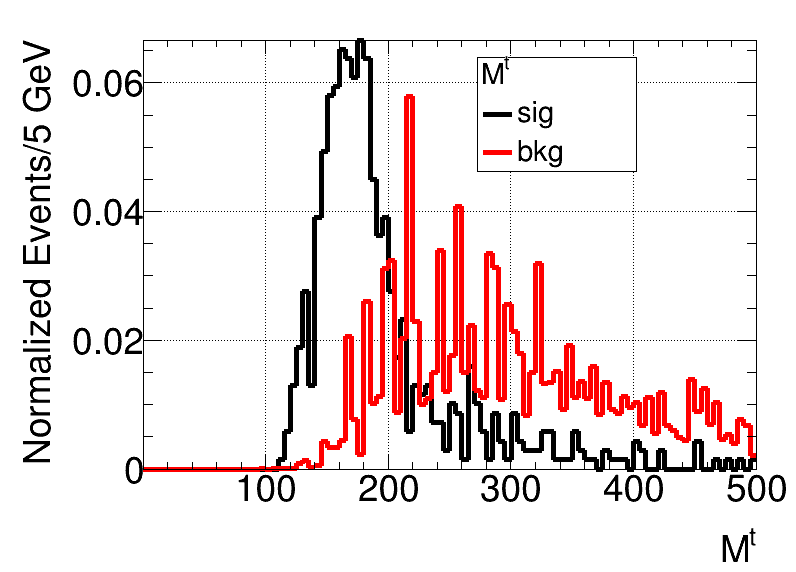} 
  \end{center}
 \end{minipage}
 \begin{minipage}{0.30\textwidth}
    \begin{center}        
      (c) Case C
 \includegraphics[height=4cm]{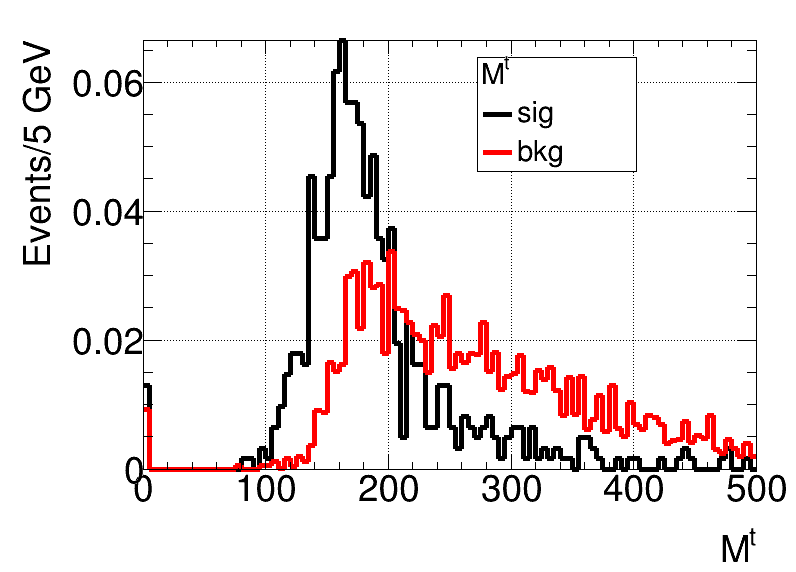} 
  \end{center}
 \end{minipage}
   \caption{The reconstructed top quark mass, $M^t$, is shown for Case A (a), Case B (b) and Case C (c). The signal events are generated by using BP1 in Tab. \ref{tab:benchmarks_points}.}\label{f_t_mass}
\end{figure} 

In a signal event, the charged Higgs boson is produced by a top quark (or antiquark) decay. Therefore, we can reconstruct the top quark mass by using the momentum of the charged Higgs boson and that of a non-$\tau$ jet. 
In our reconstruction method, we loop over all such jets and select the one which yields a reconstructed top quark mass, $M^t$, which is closest to the real one. In Fig. \ref{f_t_mass}, we present its distribution.

\item \textbf{The reconstructed $tj$ system}

\begin{figure}[ht]
  \begin{minipage}{0.30\textwidth}
    \begin{center}    
      (a) Case A    
 \includegraphics[height=4cm]{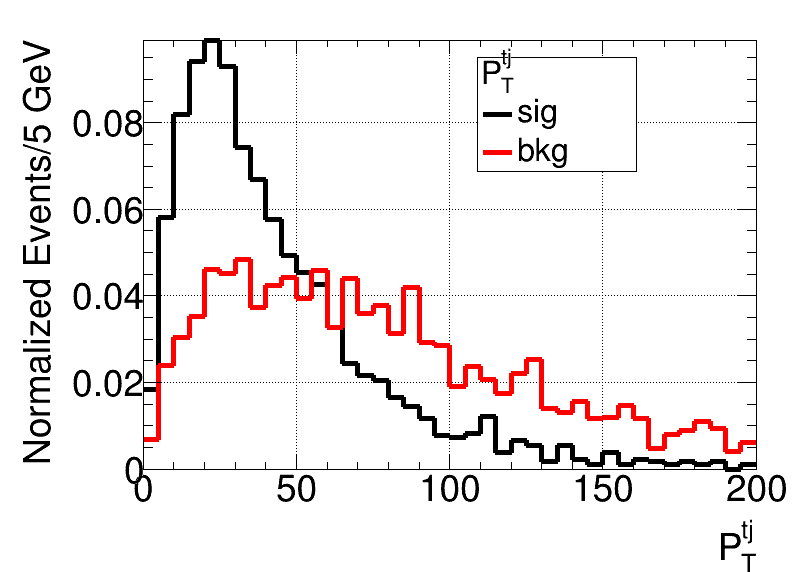} 
  \end{center}
 \end{minipage} 
   \begin{minipage}{0.30\textwidth}
    \begin{center}        
      (b) Case B
 \includegraphics[height=4cm]{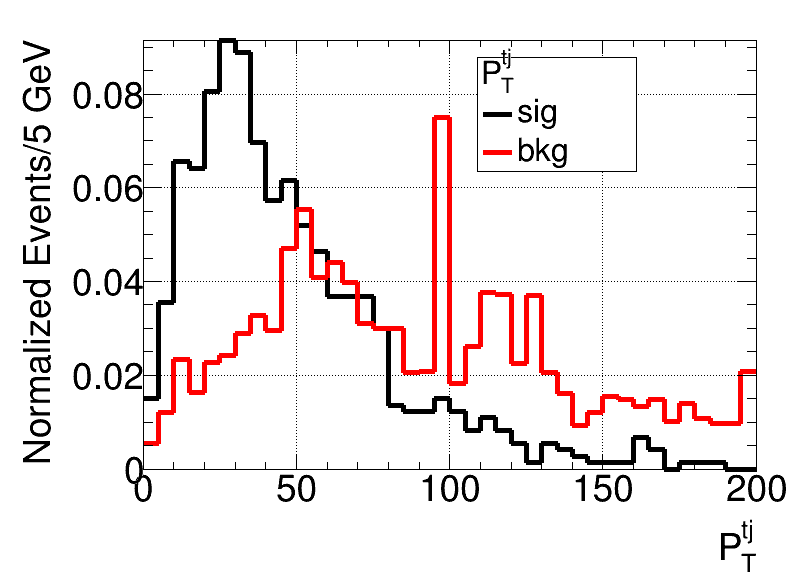} 
  \end{center}
 \end{minipage}
 \begin{minipage}{0.30\textwidth}
    \begin{center}        
      (c) Case C
 \includegraphics[height=4cm]{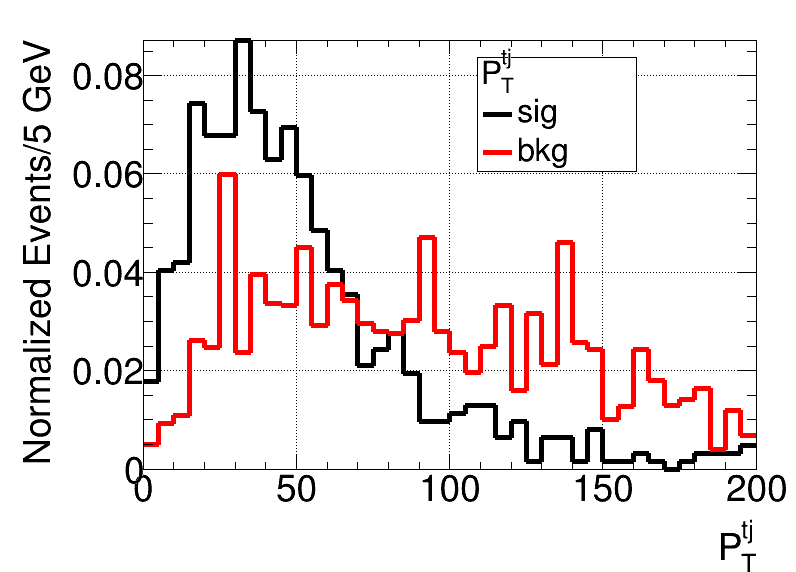} 
  \end{center}
 \end{minipage} 
  \caption{The transverse momentum distributions of the reconstructed $tj$, $P_T^{tj}$, system is shown for Case A (a), Case B (b) and Case C (c). The signal events are generated by using BP1 in Tab. \ref{tab:benchmarks_points}.}\label{f_tj_pt}
\end{figure}

When two non-$\tau$-jets are identified, as mentioned above, we pick one of these to reconstruct the top quark mass, then the other one is used to reconstruct the center-of-mass frame of the top-jet ($tj$) system, which transverse momentum, $P_{T}^{tj}$, is expected to be small, as shown in Fig.  \ref{f_tj_pt}.

\end{itemize}


 \begin{table}
 \begin{center}
 \begin{tabular}{|c| c| c| c| }
 \hline
Cuts &    Case A    & Case B & Case C\\
 \hline
 2nd jet $E_T$ & 130 GeV &  150 GeV & 200 GeV \\
  \hline
 1st lepton $E_T$ & 60 GeV & 60 GeV & 100 GeV  \\
 \hline
 2nd lepton $E_T$ & - & 50 GeV & 60 GeV  \\
  \hline
 3rd lepton $E_T$ & - & - & 35 GeV  \\
  \hline
 $M^{h^{0}}$ & [40, 100] GeV & [10, 80] GeV & [20, 75] GeV  \\
  \hline
$P_T^{h^{0}}$ & [0, 150] GeV & [0, 100] GeV & [0, 120] GeV \\
  \hline 
 $M^{H^{\pm}}$ & [80, 300] GeV & [60, 250] GeV & [80, 250] GeV \\
  \hline
 $M^{t}$ & [0, 400] GeV &  [0, 400] GeV &  [0, 400] GeV \\
  \hline
 $P_T^{tj}$ & [0, 150] GeV &   [0, 150] GeV &  [0, 150] GeV \\
  \hline
 BDTG & [0.4,1]  & [-0.6,1]  & [0.4,1]\\
\hline
 \end{tabular}
     \caption{Kinematic cuts for the analysis of BP1 in the $tj$ channel are shown. }\label{t_cut_table}
 \end{center}
\end{table}

To optimize the application of cuts upon the various kinematic observables we have just discussed and to improve the signal-to-background rate, we adopt the Gradient Boosted Decision Tree (BDTG) approach, which is one of the Multi-Variate Analysis (MVA) methods. The latter has recently been widely used in data analysis and we adopt the BDTG method implemented in the Toolkit for MVA (TMVA)  package of ROOT  \cite{Antcheva:2011zz}.
In the training stage, we have used the following input variables: all final state (standard jet, $\tau$ jet, and lepton) transverse momenta, the invariant mass and transverse momentum of the  $h^0$($A^0$) state, the reconstructed charged Higgs boson and top quark masses plus the transverse momentum of the $tj$ system. 
\begin{figure}[ht]
  \begin{minipage}{0.30\textwidth}
    \begin{center}    
      (a) Case A    
 \includegraphics[height=4cm]{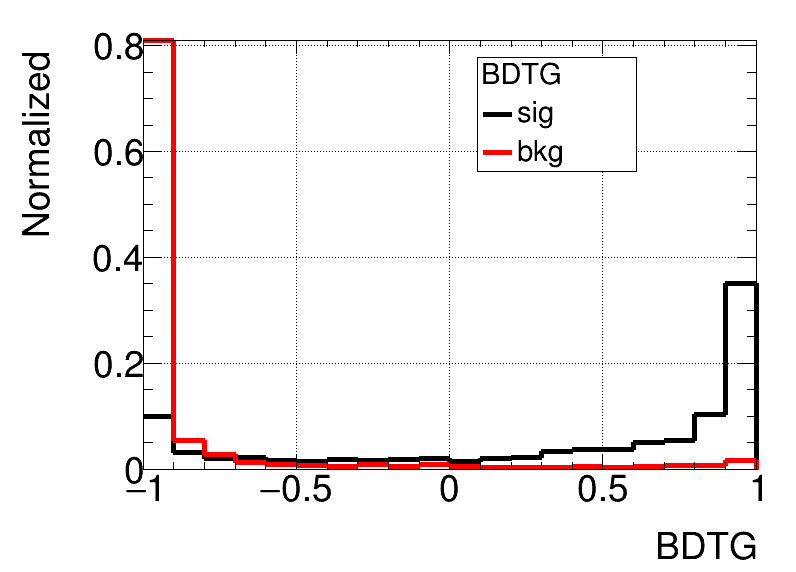} 
  \end{center}
 \end{minipage} 
   \begin{minipage}{0.30\textwidth}
    \begin{center}        
      (b) Case B
 \includegraphics[height=4cm]{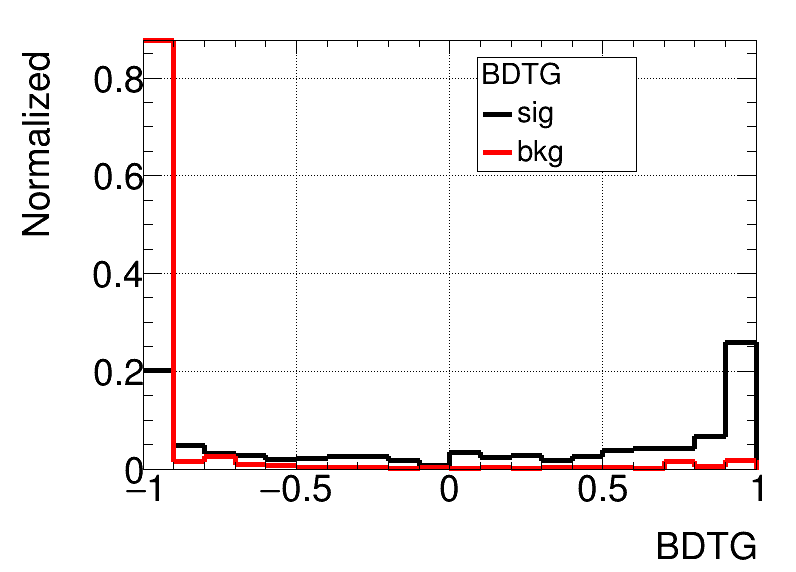} 
  \end{center}
 \end{minipage}
 \begin{minipage}{0.30\textwidth}
    \begin{center}        
      (c) Case C
 \includegraphics[height=4cm]{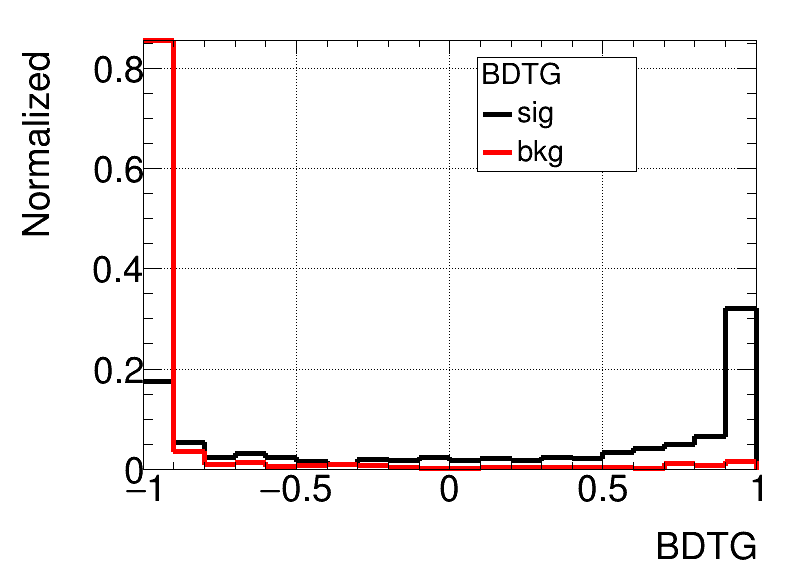} 
  \end{center}
 \end{minipage} 
  \caption{The BDTG output in the $tj$ channel analysis is shown for Case A (a), Case B (b) and Case C (c). The signal events are generated by using BP1 in Tab. \ref{tab:benchmarks_points}.}\label{f_BDTG}
\end{figure} 
The distribution of the BDTG output for both signal and background is shown in Fig. \ref{f_BDTG}, from which it is noticed that the observables listed above as input are indeed efficient to distinguish between the two samples for all three Cases considered (A, B, and C).
%
To maximally exploit the efficiency of the BDTG method, all kinematic cuts are chosen rather loose and thus only remove the most irrelevant background events. The details of the cuts used for three cases for BP1 are summarised in Tab. \ref{t_cut_table}, alongside the BDTG cut. We stress that the BDTG output is a much better variable to use to enhance the signal-to-background rate than that of a traditional cut flow method based on more stringent requirements placed upon the aforementioned (or other) kinematic variables.  

Although, for reasons of space, we have illustrated here only the response of BP1 (and associated background) to our selection, the pattern for the other BPs is very similar, the main difference being at the BTDG training stage where the input signal events are generated by the different kinematic parameters about each BP. We have also applied our analysis to the case of $\sqrt{s}=13$ TeV. Based on our selection, the final significances for all Cases A, B, and C, as well as all six BPs, are summarised in Tab. \ref{t_significance1}.  
The combined significances for the three cases are calculated too. The luminosity is assumed to be 100 $\text{fb}^{-1}$ at both 13 and 14 TeV, hence we can compare like-for-like the scope of Run 2 and Run 3 of the LHC, when the latter is only affected by the change of beam energy. For example, by looking at the combined rates at 14 TeV, it is clear that the significances of BP1 and  BP3--6 can be larger than 5, in fact, significantly so in some cases, thus warranting a potential discovery. In contrast,  BP2 can only afford one with some evidence of new physics. {{At 13 TeV, the results are very similar for each BPs, albeit somewhat smaller in comparison.}}

 \begin{table}
 \begin{center}
 \subcaption{$\sqrt{s}$=13 TeV}
 \begin{tabular}{|c| c| c| c| c| c| c| }
 \hline
 Significance &    BP1    & BP2 & BP3 &BP4 &BP5 & BP6\\
  \hline
Case A    &4.14&2.50&6.20&6.00&6.18&4.92\\
 \hline
Case B  &3.80&1.86&3.34&5.80&5.96&5.28\\
 \hline
Case C  &2.69&2.04&3.81&4.46&4.56&4.13\\
 \hline
Combined    &6.07&3.54&7.96&9.13&9.41&7.93\\
\hline
 \end{tabular}
 
 \subcaption{$\sqrt{s}$=14 TeV}
  \begin{tabular}{|c| c| c| c| c| c| c| }
   \hline
 Significance &    BP1    & BP2 & BP3 &BP4 &BP5 & BP6\\
 \hline
Case A    &4.29&2.59&6.44&6.19&6.41&5.08\\
 \hline
Case B  &3.96&1.94&3.49&6.07&6.26&5.53\\
 \hline
Case C &2.80&2.13&3.97&4.62&4.76&4.29\\
 \hline
Combined &6.30&3.66&8.28&9.45&9.79&8.22\\
\hline
 \end{tabular}

 \end{center}
      \caption{The final significances of the $tj$ channel for the  six BPs considered when the luminosity is 100 fb$^{-1}$ at both (a) $\sqrt{s}=13$ TeV and (b) $14$ TeV.} \label{t_significance1}
\end{table}

\subsection{Charged Higgs boson production from $pp \to tt$ processes}\label{ttproduction}
In this section, we extend our analysis to the $pp \to t {\bar t}$ production channel. 
Here, the dominant signal production process is $pp \to t {\bar t} \to H^{\pm}b t(\bar{t}) \to h^{0}(A^{0})l^{\pm}\nu b\bar b W^\pm \to \tau^+\tau^- l^{\pm}\nu bbqq$, where one has off-shell $W^\pm$ (leptonic) decays from $H^\pm$ transitions and on-shell $W^\pm$ (hadronic) decays from $t$ transitions. For this signature,
 we will consider the following main background processes: 1) $t\bar t$ production; 2) $W\tau^+\tau^-$ production; 3) $tt\ell\ell$ production. (Notice that we will instead ignore the following subdominant backgrounds, which inclusive cross sections are small when compared with those of the dominant ones and, further, can be easily suppressed by our selection cuts: $W^+W^-\tau^+\tau^-$,  single vector boson production processes, $W^+W^-$, $ZZ$, single top production as well as multi-jet processes from QCD.) 

Similar to the previous case, here too,  events are categorized into three cases in terms of the $\tau$ decay products. In all of these cases, we employ $b$-tagging as implemented in our detector emulator. 

\begin{itemize}

\item Case A: two hadronic $\tau$ decays, i.e., there are two  $\tau$ jets, one soft lepton (as usual, $l=e,\mu$), which is from an off-shell $W^\pm$ decay, two $b$-jets and two extra (untagged) jets in this signature.

\item Case B: one hadronic $\tau$, one leptonic $\tau$  (again, either into an electron or a muon), i.e., there are one $\tau$ jet, two leptons, two $b$-jets and two extra (untagged) jets in the final state.

\item Case C: two leptonic $\tau$ decays (via the above channels), i.e., there are three leptons, two $b$-jets and two extra (untagged) jets in this case.
\end{itemize}


Again, similarly to the analysis of the single top processes, to distinguish signal and background, we introduce the following features which are efficient in separating the former from the latter. 

\begin{itemize}
\item \textbf{The reconstructed $h^0$ and $A^0$ bosons}

The invariant mass of two tagged $\tau$ jets (Case A), the $\tau$ jet and the hardest lepton (Case B) and the two hardest leptons (Case C) are clustered together. For signal events, again, the invariant mass of this system correlates to the mass of the CP-even or CP-odd Higgs boson (i.e., $h^{0}$ and $A^{0}$, respectively).

\item \textbf{The reconstructed $W^\pm$ boson}

The mass of the hadronic $W^\pm$ boson can be computed from the invariant mass of two non-$b$-jets. In signal events, these two non-$ b$-jets are produced from hadronic $W^\pm$ decays and would naturally peak at $M_{W^\pm}$. In the case of leptonic decays of a $W^\pm$ boson, one can use instead the standard transverse mass using the softest lepton and the MET component, though for the signal it should be recalled that the gauge boson is off-shell.

\item \textbf{The reconstructed $H^\pm$ boson}

We can construct the mass of charged Higgs boson from the softest lepton, the MET and the reconstructed $h^{0}(A^{0})$, as previously explained.


\item \textbf{The reconstructed $t$ quarks}

The reconstructed $W^\pm$ and $H^{\mp}$ boson candidates given above and two $b$-jets are used further to reconstruct two top quark masses. However, contrary to the previous case of single top production, when only one top quark mass reconstruction is involved, here, to reduce the impact of  combinatorics, we perform  a $\chi^2$ fit, as follows:
\begin{equation}
\chi^2=( M_{b_{i}W} -m_t)^2+ (M_{b_{j} {H^{\pm}}} - m_t)^2\,,
\end{equation}
where  $m_t$ is taken as $173.5$ GeV. The combination which yields the minimal $\chi^2$ is chosen for each event.
\end{itemize}

\begin{figure}[ht]
  \begin{minipage}{0.30\textwidth}
    \begin{center}    
      (a) Case A    
 \includegraphics[height=4cm]{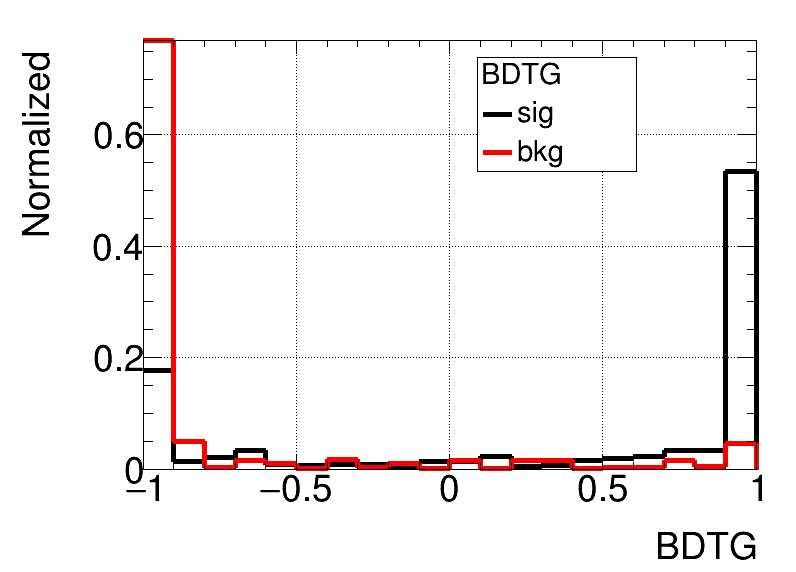} 
  \end{center}
 \end{minipage} 
   \begin{minipage}{0.30\textwidth}
    \begin{center}        
      (b) Case B
 \includegraphics[height=4cm]{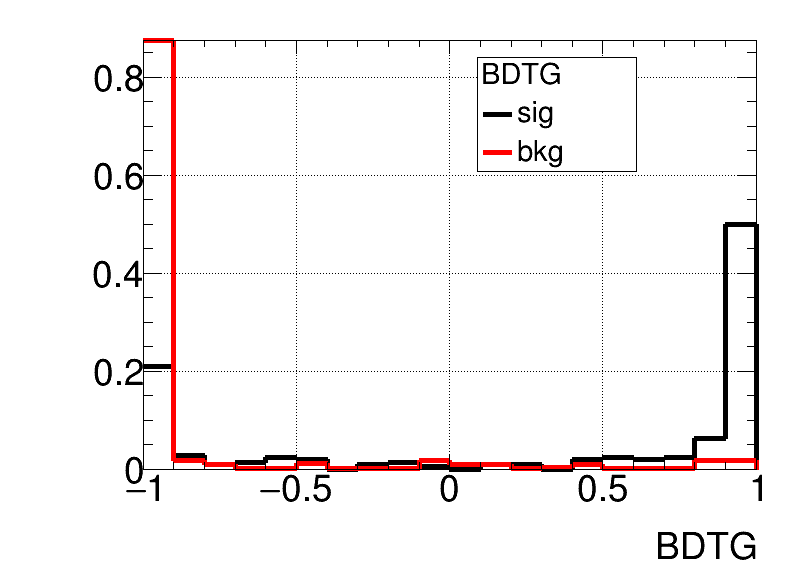} 
  \end{center}
 \end{minipage}
 \begin{minipage}{0.30\textwidth}
    \begin{center}        
      (c) Case C
 \includegraphics[height=4cm]{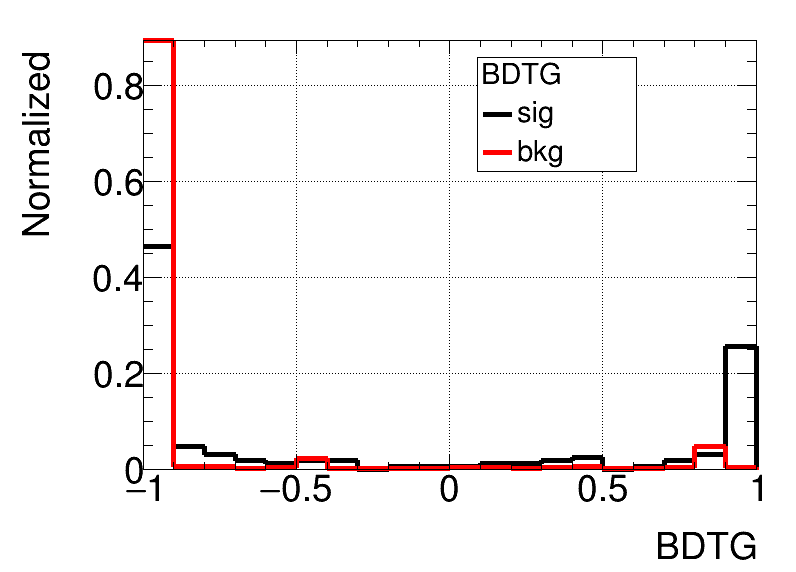} 
  \end{center}
 \end{minipage} 
  \caption{The BDTG output in the $tt$ channel analysis is shown for Case A (a), Case B (b) and Case C (c). The signal events are generated by using BP1 in Tab. \ref{tab:benchmarks_points}.}\label{f_BDTG2}
\end{figure}

In our BDTG method, at the training stage, we use here the following input variables: all final state ($b$- and non-$b$-jet, $\tau$ jet and lepton) transverse momenta, the reconstructed mass of the  $h^0$($A^0$) state, the reconstructed charged Higgs boson and top quark masses. The distribution of the BDTG output is shown in Fig. \ref{f_BDTG2}. Again, we optimize the cuts of the BDTG output to obtain the best significance for each BP, see Tab.~\ref{t_cut_table_tt}.  The significances for each of the three cases of $\tau$ final states and the combined ones for the six BPs are presented in Tab.~\ref{t_significance2}, for our two customary choices of energy and luminosity. {{At both 13 TeV and 14 TeV}}, the combination of all $\tau^+\tau^-$ decay modes can afford one with significant potential discovery for all BPs studied.

 \begin{table}
 \begin{center}
 \begin{tabular}{|c| c| c| c| }
 \hline
Cuts &    Case A    & Case B & Case C\\
 \hline
 $M^{h^{0}}$ & [0, 90] GeV & [0, 105] GeV & [20, 100] GeV  \\
  \hline
 $M^{top1}$ & [130, 220] GeV & [100, 250] GeV & [80, 250] GeV \\
  \hline
 $M^{H^{\pm}}$  & [80, 220] GeV &  [80, 220] GeV &  [80, 220] GeV \\
  \hline
 BDTG & [0.8,1]  & [-0.9,1]  & [-0.9,1]\\
\hline
 \end{tabular}
     \caption{Kinematic cuts for the analysis of BP1 in the $t\bar{t}$ channel are shown. }\label{t_cut_table_tt}
 \end{center}
\end{table}

 \begin{table}
 \begin{center}
 \subcaption{$\sqrt{s}$=13 TeV}
 \begin{tabular}{|c| c| c| c| c| c| c| }
 \hline
 Significance &    BP1    & BP2 & BP3 &BP4 &BP5 & BP6\\
  \hline
Case A    &6.10&6.32&5.36&9.94&8.77&9.04\\
 \hline
Case B  &5.65&6.22&5.37&8.26&7.11&5.56\\
 \hline
Case C &3.53&4.02&2.92&4.49&4.81&5.20\\
 \hline
Combined    &8.99&9.70&7.96&13.11&12.20&11.58\\
\hline
 \end{tabular} 

 \subcaption{$\sqrt{s}$=14 TeV} 
 \begin{tabular}{|c| c| c| c| c| c| c| }
 \hline
 significance &    BP1    & BP2 & BP3 &BP4 &BP5 & BP6\\
 \hline
Case A    &6.38&6.62&5.58&10.39&9.17&9.42\\
 \hline
Case B  &5.93&6.54&5.65&8.65&7.46&5.88\\
 \hline
Case C  &3.70&4.22&3.06&4.71&5.02&5.44\\
 \hline
combine    &9.42&10.18&8.32&13.73&12.75&12.14\\
\hline
 \end{tabular}

 \end{center}
     \caption{The final significances of the $tt$ channel for the  six BPs considered when the luminosity is 100 fb$^{-1}$ at both (a) $\sqrt{s}=13$ TeV and (b) $14$ TeV.}\label{t_significance2}
\end{table}

\section{Conclusions}

In summary, in this paper, building upon previous work of some of us, we have proven that bosonic decays of light charged Higgs boson, i.e., $H^\pm \to A^0 W^{\pm(*)}$ or $h^0 W^{\pm(*)}$, where the charged Higgs boson is produced in a top (anti)quark decay, 
can be accessed already during Run 2 of the LHC and certainly at Run 3. This can be done in the $\tau^+\tau^-$ decay channel of the two aforementioned neutral Higgs bosons so long that either or both of these are lighter than the SM-like Higgs boson discovered at the CERN collider back in 2012. This spectrum configuration is available in a 2HDM Type-I, wherein it is possible to identify the latter state with the heaviest CP-even Higgs boson of the model, i.e., $H^0$. This can be achieved in the presence of all theoretical and experimental constraints presently available so that this BSM scenario
is a prime candidate for a detectable new physics signal at the LHC. This requires to exploit all possible production modes of the top (anti)quark, i.e., 
both single and double top (anti)quark production,  and all possible $\tau^+\tau^-$ decays modes, i.e., fully hadronic, fully leptonic and semi-leptonic (or semi-hadronic).

We have come to the above conclusions following a complete scan of the parameter space of the 2HDM Type-I and a sophisticated MC analysis of several BPs maximizing, in turn, the $A^0 W^{\pm(*)}$ or $h^0 W^{\pm(*)}$ decay rates of the $H^\pm$ state. For both the former and the latter we have used sophisticated numerical tools so that we are confident of the solidity of our results. We look forward to the ATLAS and/or CMS collaborations to test our BSM scenarios through the advocated signatures, which may well serve the purpose of offering additional discovery modes of not only a light charged Higgs boson state but also one neutral Higgs boson state lighter than the discovered SM-like one. Unfortunately, not both $A^0$ and $h^0$ states can be accessed at the same time in this way, as the $H^\pm\to A^0 W^{\pm(*)}$ decay rate is largest when the     $H^\pm\to h^0 W^{\pm(*)}$ one is smallest (and vice versa). 

\section*{Acknowledgments}
This work is supported by the Moroccan Ministry of Higher Education and Scientific Research MESRSFC and
CNRST: Projet PPR/2015/6.
SM is supported also in part through the NExT Institute and the STFC consolidated Grant No. ST/L000296/1. AA , RB and SM acknowledge the H2020-MSCA-RISE-2014 Grant No. 645722 (NonMinimalHiggs). Q.S. Yan is supported by the Natural Science Foundation of China under grants No. 11475180 and No. 11875260.

 \end{document}